\documentclass[journal]{IEEEtran}
\IEEEoverridecommandlockouts
\usepackage{epsfig,times}
\usepackage{subfigure}
\usepackage{multirow}
\usepackage{epsfig}
\usepackage{graphicx}
\usepackage{url}
\usepackage{times}
\usepackage{subfigure}
\usepackage{acronym}
\usepackage{psfrag}
\usepackage{booktabs}
\usepackage{amsmath}
\usepackage{array}
\usepackage{amssymb}
\usepackage{cite}
\usepackage{epsfig,times}
\usepackage{subfigure}
\usepackage{multirow}
\usepackage{epsfig}
\usepackage{graphicx}
\usepackage{url}
\usepackage{times}
\usepackage{subfigure}
\usepackage{acronym}
\usepackage{psfrag}
\usepackage{booktabs}
\usepackage{amsmath}
\usepackage{array}
\usepackage{amssymb}
\usepackage{fontenc}
\usepackage{booktabs}
\usepackage{color}
\usepackage{hhline}
\usepackage{soul}
\usepackage{tabu}
\usepackage{makecell} 
\usepackage{xcolor}

\begin{document}

\title{On-CMOS High-Throughput Multi-Modal Amperometric DNA Analysis with Distributed Thermal Regulation}

\author{\IEEEauthorblockN{Hamed M.~Jafari,~\IEEEmembership{Member,~IEEE,}
        Xilin~Liu,~\IEEEmembership{Senior Member,~IEEE,}
        and~Roman~Genov,~\IEEEmembership{Senior Member,~IEEE}}
        
        \vspace{-0.8cm}
        
}


\maketitle

\begin{abstract}

Accurate temperature regulation is critical for amperometric DNA analysis to achieve high fidelity, reliability, and throughput. In this work, a 9$\times$6 cell array of mixed-signal CMOS distributed temperature regulators for on-CMOS multi-modal amperometric DNA analysis is presented. Three DNA analysis methods are supported, including constant potential amperometry (CPA), cyclic voltammetry (CV), and impedance spectroscopy (IS). In-cell heating and temperature sensing elements are implemented in standard CMOS technology without post-processing. Using proportional-integral-derivative (PID) control, the local temperature can be regulated to within ±0.5$^{\circ}$C of any desired value between 20$^{\circ}$C and 90$^{\circ}$C. The two computationally intensive operations in the PID algorithm, multiplication, and subtraction, are performed by an in-cell dual-slope multiplying ADC in the mixed-signal domain, resulting in a small area and low power consumption. Over 95\% of the circuit blocks are synergistically shared among the four operating modes, including CPA, CV, IS, and the proposed temperature regulation mode. A 3mm$\times$3mm CMOS prototype fabricated in a 0.13$\mu$m CMOS technology has been fully experimentally characterized. Each channel occupies an area of 0.06mm$^{2}$ and consumes 42$\mu$W from a 1.2V supply. The proposed distributed temperature regulation design and the mixed-signal PID implementation can be applied to a wide range of sensory and other applications.

\end{abstract}


\IEEEpeerreviewmaketitle


\section{Introduction}

\begin{figure}[!ht]
\centering
\includegraphics[width=0.45\textwidth]{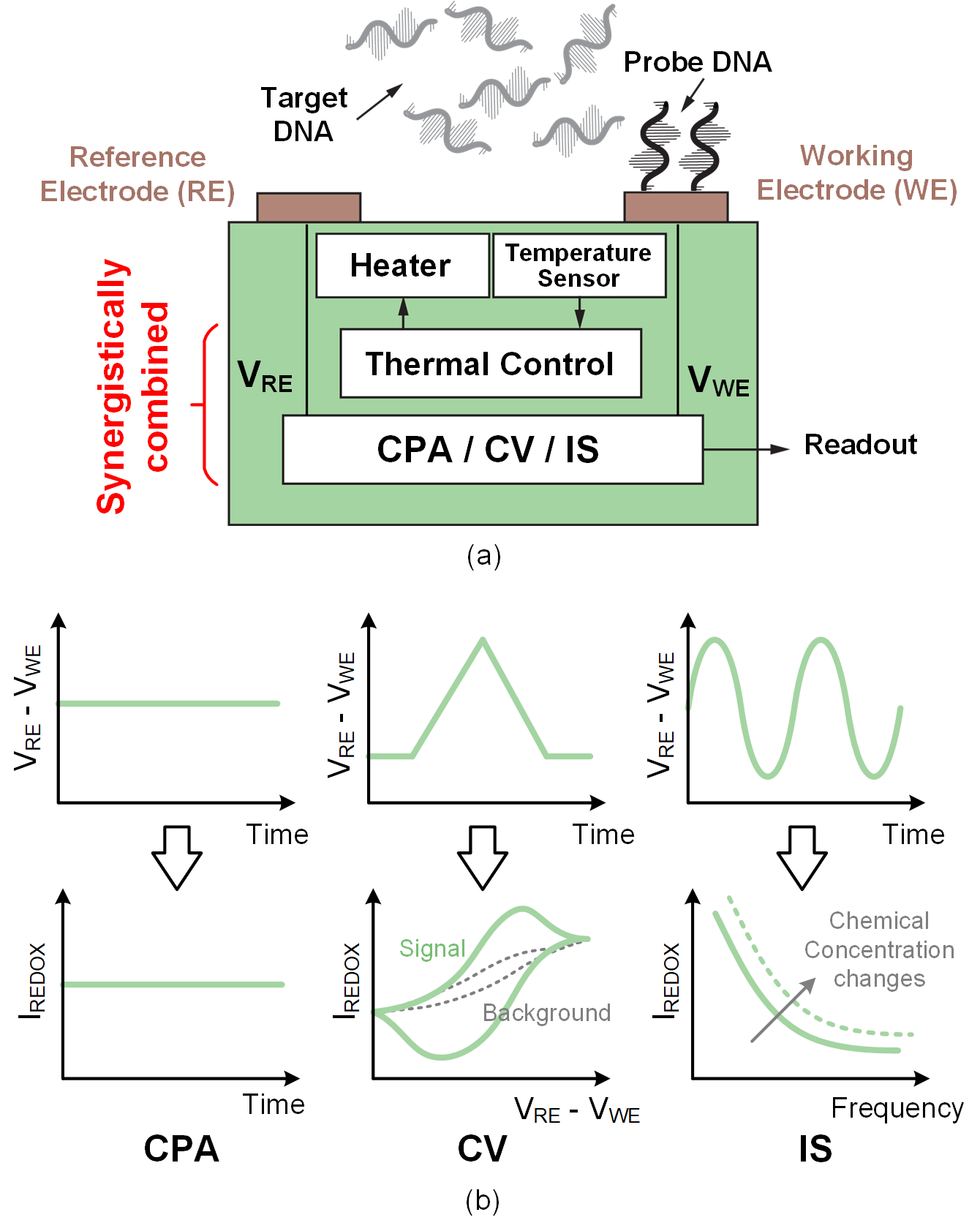}
\caption{(a) Conceptual block diagram of one of the temperature-regulated amperometric DNA analysis channels. (b) Illustration of the basic operations of the three amperometric DNA analysis modes supported in this work: constant potential amperometry (CPA), cyclic voltammetry (CV), and impedance spectroscopy (IS).}
\label{fig:intro}
\end{figure}

Conventional optical DNA analysis methods typically require sample flushing, thus cannot be implemented in real-time~\cite{Jang}. Amperometric electrochemical DNA analysis supports real-time operation \cite{Levine}. Recently, several CMOS system-on-chip (SoC) designs for amperometric electrochemical DNA analysis have been developed \cite{Heer,Wang,Garner,Zhang2020,Manickam2017,Manickam2019,Park2020,Tedjo2020}. On-CMOS DNA analysis permits advantageous high-density, low-noise, low-cost monolithic integration of various modules for sensing, signal processing, communication, and power management \cite{JNE2021}.
Fig. \ref{fig:intro} illustrates the operating principles of amperometric DNA analysis. First, the surface of the working electrode (WE) is functionalized with probe DNA \cite{Heer}. Then, a voltage is applied between the reference electrode (RE) and the WE. The voltage waveform is set to be a constant voltage for constant potential amperometry (CPA), a bidirectional ramp for cyclic voltammetry (CV), and a sinusoid for impedance spectroscopy (IS).
The hybridization of the probe DNA with the target DNA alters the surface properties of the WE, such as the impedance or the surface charges. This variation will result in a waveform change of the redox current that indicates the kinetics and thermodynamics of the chemical reactions \cite{Drummond}.

Temperature is a key constraint in all DNA analysis methods. Studies have shown that DNA hybridization is highly dependent on the temperature \cite{Sassolas,Liu,Malorny}. For a temperature increase of 10$^{\circ}$C, the redox current will drop approximately by 10\% \cite{Sassolas}. Additionally, time-varying temperature control in the range of 20$^{\circ}$C to 90$^{\circ}$C is required for DNA multiplication through polymerase chain reaction (PCR). Therefore, a temperature regulation system is of great benefit for DNA sensing applications. 
In practice, off-array CMOS processing modules often cause temperature variation across the DNA analysis array, and do not scale well to large sensing arrays. Thus, distributed in-channel temperature sensing and regulation are necessary. In addition, a channel-level thermal regulation scheme permits the generation of spatial temperature gradients for advanced non-uniform DNA analysis \cite{Harper2007, Kopparthy2020}.

\begin{figure*}[!ht]
   \centering
  \leavevmode
	\includegraphics[width=0.85\textwidth]{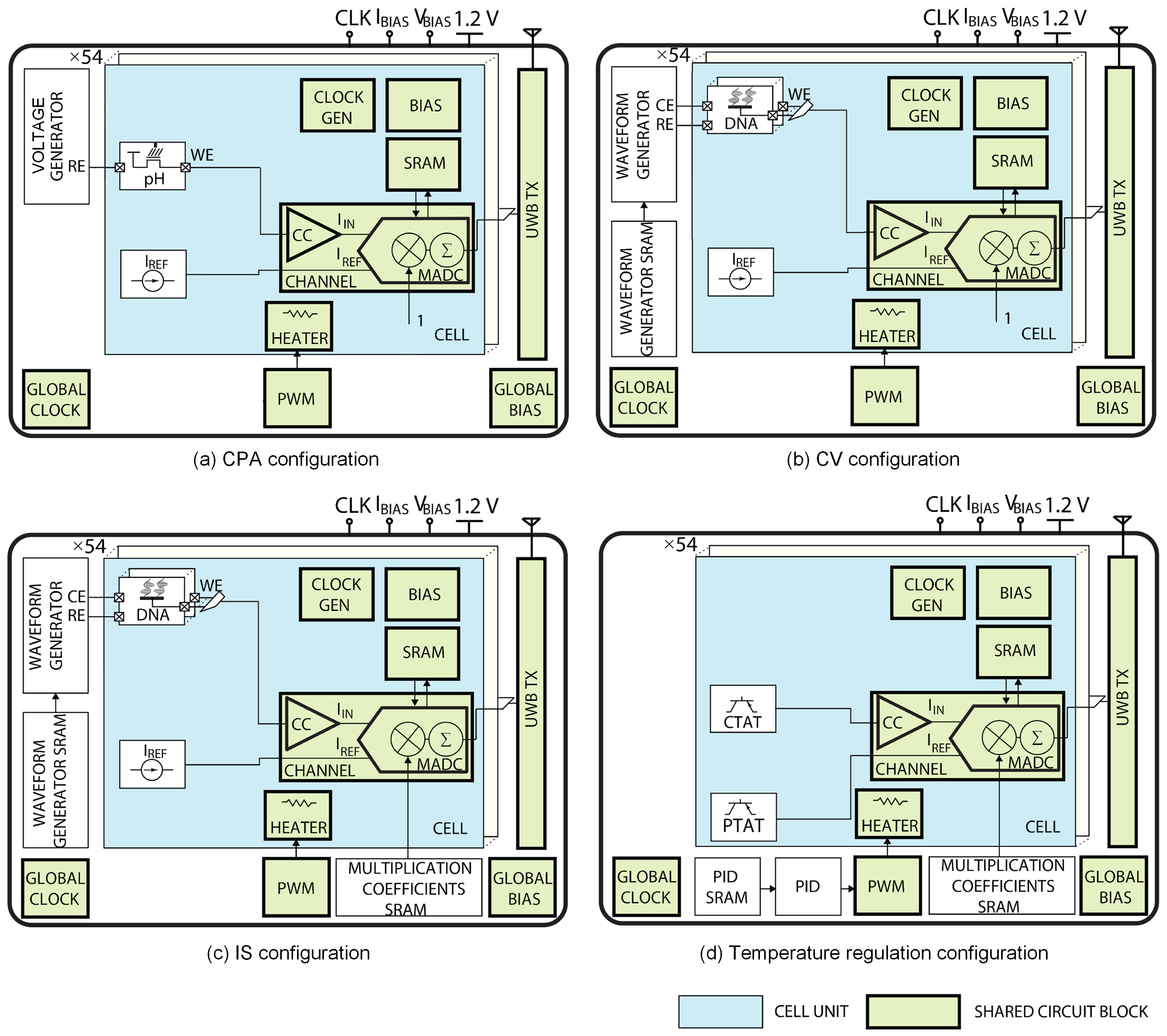}
   \caption{Four modes of operation of the proposed multi-modal DNA analysis SoC. (a) CPA configuration \cite{VLSI}, (b) CV configuration \cite{JSSC}, (c) IS configuration \cite{TBIOCAS}, and (d) temperature regulation configuration (the focus of this work). The blocks with a thick outline and highlighted in light green are synergistically shared in all modes, with over 95\% of all area reused in each operating mode.}
   \label{fig:sys_overview}
\end{figure*}


Several temperature regulation designs for chemical sensing applications have been reported \cite{Christen,Mason_T,Wang,frey,Manickam2019,Park2020,Tedjo2020}. These designs typically use resistive heating and sensing elements. Paper \cite{Christen} presents a temperature sensing and regulation microsystem for biological applications that achieves a single fixed temperature of 37$^{\circ}$C on the backside of the chip, with an off-chip digital proportional-integral-derivative (PID) controller. Paper \cite{Mason_T} presents a 3$\times$3 micro-hotplate array for a protein-sensing application. In this design, the temperature of the chip is regulated up to 45$^{\circ}$C with an off-chip digital PID controller. Paper \cite{Wang} presents a 16-channel, frequency-shift CMOS magnetic biosensor array for a protein-sensing application. This design consists of an array of in-channel heaters, with a single temperature sensor and an off-chip analog PID controller. Paper \cite{frey} presents a digital CMOS micro-hotplate array architecture for a gas-sensing application. The array consists of 3$\times$3 gas-sensing sites. Each site consists of a heater, a temperature sensor, and a delta-sigma modulated PID controller. The micro-hotplate achieves an operating temperature of up to 350$^{\circ}$C. A 40-channel DNA analysis SoC based on ISFET sensors for rapid point-of-care DNA detection is reported in \cite{Garner}. This design includes in-channel heaters, temperature sensing elements, and a dedicated ADC. A fully digital on-chip PID controller is utilized for temperature regulation. In summary, most existing closed-loop temperature regulation microsystems use off-chip controllers, with few exceptions \cite{Garner,Park2020}. Conventional on-chip controller designs often require a large silicon area along with high power consumption, which is prohibitive for most on-chip electrochemical sensing applications.

In this work, we present a compact, distributed temperature regulator design implemented in 0.13$\mu$m standard CMOS technology. 
Fig.~\ref{fig:intro} depicts the block diagram of the proposed temperature-regulated electrochemical cell. It consists of a temperature sensor, a heater, a closed-loop regulator, a working electrode, a reference electrode, and a potentiostat. 
We chose the PID algorithm for the closed-loop regulation because of its high efficacy in temperature control systems \cite{TEMP1, TEMP2}. PID controllers have been implemented on-chip in analog \cite{PID_TBioCAS, PID_BioCAS} and digital domains \cite{PID_digital}. Here we introduced a novel mixed-signal design that yields a compact integration with low power dissipation. This design permits distributed temperature regulation in large-scale parallel sensing applications.

The presented temperature controller is a part of a highly-integrated multi-functional SoC that performs on-chip amperometric electrochemical DNA analysis. The overall SoC consists of 54 sensing channels with 600 gold microelectrodes. It performs label-free DNA analysis and pH sensing for prostate cancer detection. During the design, we synergistically share the circuits among different operation modes to achieve high integration and low resource overhead. We have previously reported on the cyclic voltammetry DNA sensing modality in \cite{JSSC}. We briefly discussed how the DNA sensing and the temperature control circuits are integrated in \cite{VLSI}. The impedance modality was introduced by us in \cite{TBIOCAS}.
In this paper, we presented a unified review of how all of these sensing modalities are implemented on the same IC with minimal overhead, and focus on presenting the details of the distributed mixed-signal temperature regulator, including the system architecture, algorithm design, circuit implementation, and experimental results. To the best of our knowledge, this work presents the first distributed temperature regulation design with on-chip PID control for DNA analysis. This design permits an accurate thermal control for compensating the temperature variation across the DNA analysis array while potentially enabling advanced temperature gradient generation. We developed a new way of implementing PID control using mixed-signal circuits (e.g., multiplying ADC), which avoids the computationally intensive multiplications of the PID algorithm in the digital domain, rendering compact silicon area and low power consumption. The proposed design is required for the three sensory modalities: CPA, CA, and IS, and reuses over 95\% of their circuit implementation. It can be used in a wide range of applications beyond DNA analysis, including lab-on-chip designs, in-vitro systems, and implantable medical devices (e.g., retinal prosthesis array).



The rest of the paper is organized as follows. Section II overviews the multi-modal DNA analysis SoC. Section III provides background on temperature regulation principles. Section IV presents the VLSI architecture of the temperature controller. Section V details the circuit implementation. Section VI shows the experimental results obtained from the fabricated 0.13$\mu$m CMOS prototype. Finally, Section VII concludes the paper.

\section{Multi-modal DNA Analysis SoC with Synergistic Circuit Sharing}

Fig. \ref{fig:sys_overview} shows an overview of the presented multi-modal DNA analysis SoC, which can be configured to perform CPA (Fig.~\ref{fig:sys_overview}(a) \cite{VLSI}), CV (Fig.~\ref{fig:sys_overview}(b) \cite{JSSC}), IS (Fig.~\ref{fig:sys_overview}(c) \cite{TBIOCAS}), and temperature regulation~(Fig. \ref{fig:sys_overview}(d)).


The SoC includes a 9$\times$6 array of low-power low-noise sensory circuit cells, a programmable waveform generator, on-chip SRAM memories, a shared PID controller, a digital PWM, an all-digital ultra-wideband (UWB) transmitter, and on-chip biasing and clock generation circuits. Each sensory cell includes a current-to-digital conversion channel, an array of DNA sensors, a pH sensor, several BJTs to generate the CTAT and PTAT currents, and a heater for temperature regulation. Each cell also integrates a bias voltage generator, clock generation circuitry, and SRAM memory that can be programmed individually.

Each current-to-digital conversion channel consists of either a transimpedance amplifier (TIA) or a current conveyor (CC) and a dual-slope multiplying ADC (MADC). The dual-slope MADC multiplies the sensor's response with a set of digital coefficients, and outputs the corresponding digital word. The digital output of each channel is serialized on the chip and then wirelessly transmitted using the all-digital UWB transmitter~\cite{JSSC}.

The SoC reuses a number of blocks (denoted by bold outlines and light green color in Fig.~\ref{fig:sys_overview}) in different configurations. The circuit sharing design yields a significant saving in the integration area. Specifically, the circuits that occupy 95\% of the in-cell area are shared with other modes of operation. For completeness, all key functionalities of the SoC are summarized in this section. For the detailed implementation of the CPA, CV and IS, please refer to our previous publications in \cite{VLSI}, \cite{JSSC} and \cite{TBIOCAS}, respectively. 

\begin{figure*}[!ht]
	\centering
	\includegraphics[width=1\textwidth]{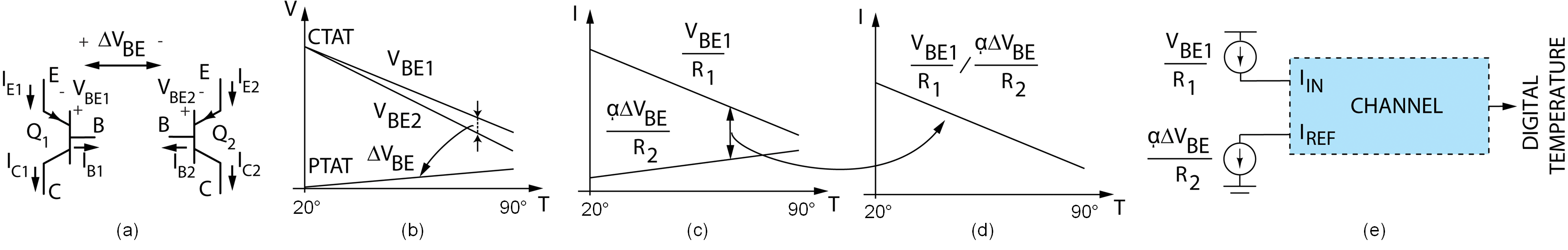}
	\caption{(a) A pair of BJTs for generating \textit{$\Delta$V$_{BE}$} and \textit{V$_{BE}$}, (b) temperature dependence of the base-emitter voltage, (c) generation of PTAT and CTAT current, (d) conceptual plot of the proposed temperature dependent current generation, and (e) utilization of a current-to-digital channel for temperature measurements.}
	\label{fig:temp_operation}
\end{figure*}

\subsection{Constant Potential Amperometry (CPA) Mode}
Fig.~\ref{fig:sys_overview}(a) shows the functional blocks for pH sensing using CPA analysis. Sensing pH is required as a part of DNA analysis. The in-channel pH sensor is implemented by a floating gate PFET, with the gate connected to the top metal layer for forming a floating gate electrode. The CMOS passivation layer is used as the pH-sensitive membrane, which gives a linear pH response with a sensitivity of about 56mV/pH, depending on the stoichiometry. The pH sensors are directly interfaced to the current recording channels, with gate voltage set by the on-chip reference electrode. In this configuration, both Vgs and Vds of the pH sensors are fixed. Changes in the pH level will affect the PMOS threshold voltage, resulting in the changes in the drain current, which is digitized by the recording channel. The chip temperature needs to be held at a constant value within ±0.5$^{\circ}$C.

\subsection{Cyclic Voltammetry (CV) Mode}

Fig.~\ref{fig:sys_overview}(b) shows the functional blocks for CV analysis. In this mode, each channel is multiplexed among a bank of in-cell DNA sensors. The sensors are interrogated by the on-chip programmable waveform generator that is shared among all cells. The CC front-end is utilized in this configuration. The digital data representing the stimulation waveform properties are stored in the on-chip SRAM. The current-to-digital channel quantizes the input current and outputs the corresponding digital word. No computation is required in this case and the MADC digital multiplication coefficients are set to one. The chip temperature is held at a constant value within ±0.5$^{\circ}$C.

\subsection{Impedance Spectroscopy (IS) Mode}

Fig.~\ref{fig:sys_overview}(c) shows the functional blocks for IS analysis. In this mode, a frequency response analysis (FRA) algorithm is used to extract the real and imaginary components of the biosensor impedance \cite{TBIOCAS}. The computationally intensive operations required by the FRA algorithm are performed by the in-channel MADC in the mixed-signal domain. The waveform generator produces a variable frequency sinusoidal interrogation waveform for driving the reference electrode, and front-end TIA acquires the DNA sensor's response. The MADC multiplies the sensor's response with a set of digital FRA algorithm coefficients (stored in the multiplication coefficient SRAM). Next, the MADC accumulates the results over one period of the interrogation signal (integration) for extracting the impedance. In this mode, the cell temperature is also held constant within ±0.5$^{\circ}$C.

\subsection{Temperature Regulation Mode}
It is clear that each of the three sensory modes requires accurate temperature control. Thus, the temperature regulation function is the main focus of this paper. Fig.~\ref{fig:sys_overview}(d) shows the functional blocks. In this mode, the SoC utilizes the current-to-digital channels to measure temperature. The on-chip PID controller is used to regulate each cell temperature. On-chip SRAMs are utilized to store PID coefficients and multiplication coefficients. The channel measures temperature by taking the ratio of a CTAT to PTAT currents as described in Section II. The measured temperature is fed to the on-chip PID controller. The PID controller regulates the 2D chip temperature profile by modulating the in-cell heaters. The multiplication and subtraction operations required by the PID algorithm are performed by the in-channel MADC in the mixed-signal domain, yielding a small silicon area and low power consumption.

\section{Temperature Control Principles}

\subsection{Temperature regulation}

\begin{figure}[!ht]
\centering
\includegraphics[width=8cm]{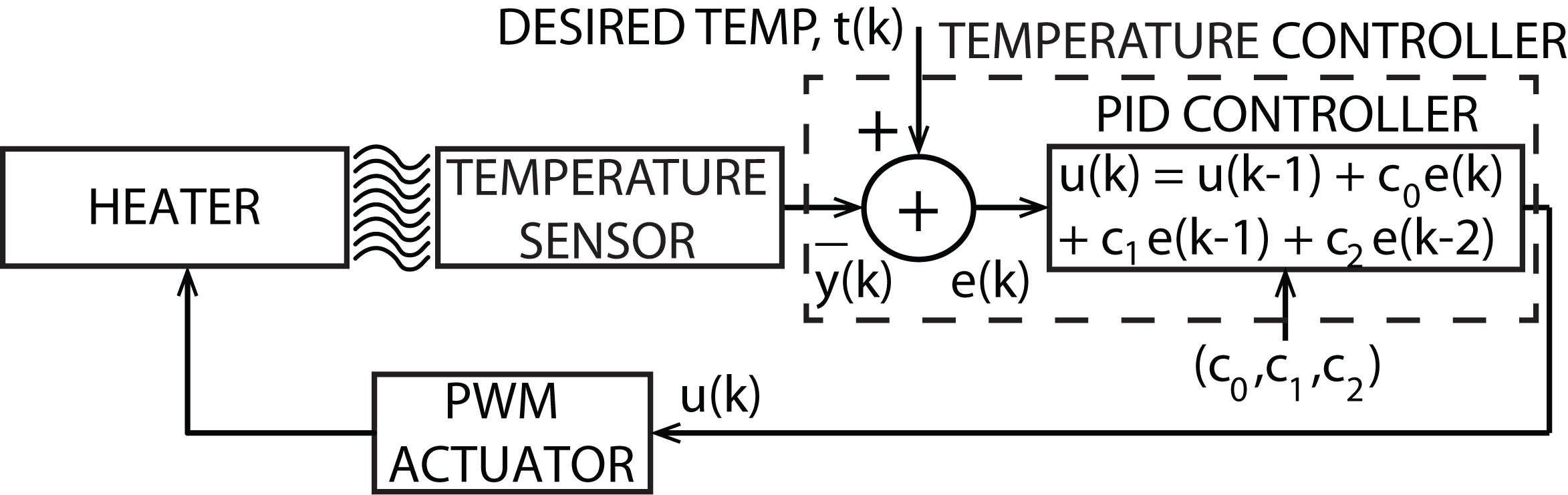}
\caption{Block diagram of the PID control of the temperature regulation loop.}
\label{fig:FIG3}
\end{figure}

The block diagram of the temperature regulation loop is shown in~Fig. \ref{fig:FIG3}. It consists of a heater, a temperature sensor, a PID controller, and an actuator. The actuator uses digital pulse-width modulation (PWM) for controlling the heater. 

The continuous-time representation of a PID controller is given by
\begin{eqnarray}
\centering
u(t)=K_{p}e(t)+K_{i}\displaystyle\int^t_0 e(t)dt+K_{d}\frac{d}{dt}e(t)
\label{eq3}
\end{eqnarray}
where \textit{K$_{p}$} is the proportional gain, \textit{K$_{i}$} is the integral gain, \textit{K$_{d}$} is the derivative gain, \textit{e(t)} is the error value, and \textit{u(t)} is the actuation value.
A discrete-time representation of~(\ref{eq3}) in Z-domain is given by
\begin{eqnarray}
\centering
D(z)=\frac{u(z)}{e(z)}=K_{p}+K_{i}\frac{T_{s}}{2}\frac{z+1}{z-1}+\frac{K_{d}}{T_{s}}\frac{z-1}{z}
\label{eq4}
\end{eqnarray}
where \textit{T$_{s}$} is the sampling period. Eq.~(\ref{eq4}) can be implemented numerically as follows
\begin{eqnarray}
\centering
   \vspace{0.005in}
u(k) = u(k-1)+c_{0}e(k)+c_{1}e(k-1)+c_{2}e(k-2)
\label{eq5}
\end{eqnarray}
where \textit{k} is the discrete time index, and \textit{c$_{0}$}, \textit{c$_{1}$}, and \textit{c$_{2}$} are the PID control coefficients.

During the operation, the PID controller calculates the error value \textit{e(k)} and attempts to minimize it by adjusting \textit{u(k)}, which sets the PWM duty cycle of the actuator. 

\begin{figure*}[!ht]
	\centering
	\includegraphics[width=0.9\textwidth]{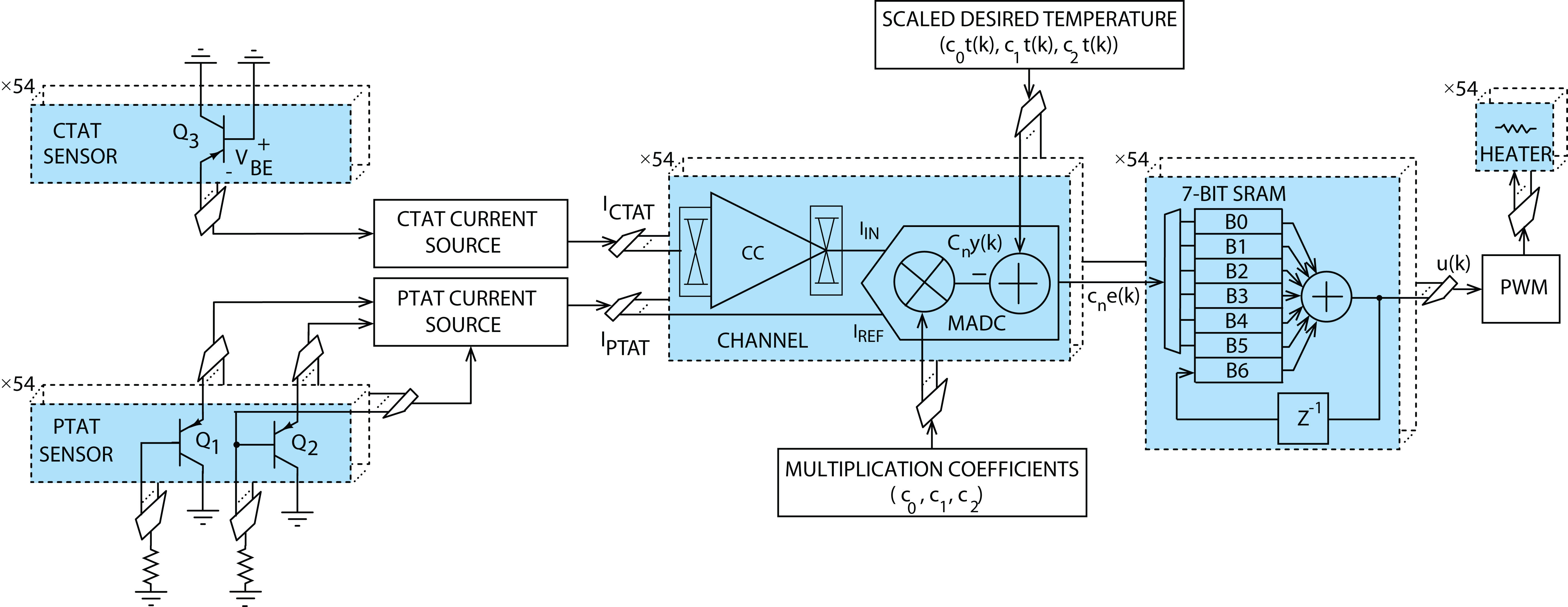}
	\caption{Top-level VLSI architecture of temperature regulation loop (shaded boxes are in-cell).}
	\label{fig:VLSI_LOOP}
\end{figure*}

\subsection{Temperature sensing}

Temperature can be measured by taking the ratio of two signals that are linearly dependent on it. In CMOS technologies, these signals can be derived from the base-emitter voltage of substrate pnp transistors \cite{Pertijs2005Feb}. These transistors use the p-substrate as the collector, the p$^{+}$-diffusion as the emitter, and an N-well as the base. Fig. \ref{fig:temp_operation}(a) shows the generation of two linearly temperature dependent signals. One is generated from the base-emitter voltage (\textit{V$_{BE}$}) of a single pnp transistor, the other is generated from the difference between the base-emitter voltages (\textit{$\Delta$V$_{BE}$}) of two pnp transistors biased at different collector current levels.

The temperature dependence of the \textit{V$_{BE}$} can be derived from the exponential relation between the collector current \textit{I$_{C}$} and the base-emitter voltage as follows.
\begin{eqnarray}
	\centering
    V_{BE}(T) &=& V_{g0}(1-\frac{T}{T_{r}})+\frac{T}{T_{r}}V_{BE}T_{r} \nonumber \\ &-&n\frac{KT}{q}ln\frac{T}{T_{r}}+\frac{KT}{q}ln\frac{I_{C}T}{I_{C}T_{r}}
\label{eq1}
\end{eqnarray}
where \textit{V$_{g0}$} is the extrapolated bandgap voltage at 0 Kelvin, \textit{n} is a process-dependent constant, \textit{K} is Boltzmann constant, \textit{q} is the electron charge,  and \textit{T$_{r}$} is an arbitrary reference temperature~\cite{T1}. As illustrated in~Fig. \ref{fig:temp_operation}(b), \textit{V$_{BE}$} is complementary to absolute temperature (CTAT) with a typical slope of~-~2~mV/K. The nonlinearity is represented by the last two terms of~(\ref{eq1}), which is negligible in the temperature range of 20$^\circ C$ to 90$^\circ C$ required by DNA analysis.

The slope of the base-emitter voltage depends on the absolute value of the collector current. This dependence can be used to generate a voltage that is proportional to absolute temperature (PTAT). The difference between the base-emitter voltages of two pnp transistors biased at two different collector currents can be expressed as follows
\begin{eqnarray}
	\centering
\Delta V_{BE}(T) &=& V_{BE2}(T)-V_{BE1}(T) \nonumber \\ &=& \frac{KT}{q}ln(\frac{I_{C2}}{I_{C1}})
\label{eq2}
\end{eqnarray}
where \textit{I$_{C1}$} and \textit{I$_{C2}$} are the collector currents of the two pnp transistors. 

Conventional temperature sensors measure the ratio of the PTAT signal to a temperature-independent reference signal. The temperature-independent reference signal can be generated by adding \textit{V$_{BE}$} to a scaled version of \textit{$\Delta$V$_{BE}$}~\cite{T1,T2,Pertijs2005Dec, Pertijs2005Feb,Liu2014}. The scaling factor $\alpha$ is used to match the temperature dependence (with an opposite sign) of \textit{V$_{BE}$} and \textit{$\Delta$V$_{BE}$}. However, the generation of the temperature-independent reference signal adds complexity to the circuit implementation, which is not suitable for a channel-level implementation. Although distributing a global temperature-independent signal is feasible, it places challenges in the signal routing if a voltage signal is used, and requires accurate matching or time multiplexing if a current signal is used. In this work, we eliminate these issues by directly taking the ratio of PTAT and CTAT signals, which have an approximately linear relationship with temperature within the range of 20$^{\circ}$C to 90$^{\circ}$C. 
The PTAT and CTAT voltages are converted to currents using resistors \textit{R$_{1}$} and \textit{R$_{2}$}. The \textit{$\alpha\Delta$V$_{BE}$/R$_{2}$} is used as the reference the dual-slope MADC and the \textit{V$_{BE}$/R$_{1}$} is used as the input signal, as illustrated in Fig. \ref{fig:temp_operation} (e). As a result, the dual-slope MADC inherently digitizes the ratio of \textit{$\alpha\Delta$V$_{BE}$/R$_{2}$} to \textit{V$_{BE}$/R$_{1}$} \cite{JSSC}. The digitized output number represents the temperature. The factor $\alpha$, the resistance \textit{R$_{1}$} and \textit{R$_{2}$} are set such that both the input and the reference currents utilize the full dynamic range of the channel over the operating temperature range.



\section{VLSI Architecture}

The top-level VLSI architecture of the temperature regulation loop is shown in Fig. \ref{fig:VLSI_LOOP}. The regulation loop consists of CTAT and PTAT BJTs, PTAT and CTAT current sources, a current-to-digital channel, a 7-bit SRAM bank with an adder, a 12-bit digital PWM, and an in-cell heater. The in-cell BJTs are interfaced sequentially to the current sources that generate \textit{I$_{PTAT}$} and \textit{I$_{CTAT}$}. \textit{I$_{CTAT}$} is used as the input to the channel, and \textit{I$_{PTAT}$} is used as the reference current. The channel determines the ratio of \textit{I$_{CTAT}$} to \textit{I$_{PTAT}$}. Both currents are scaled such that their magnitude fits within the channel dynamic range over the operating temperature range. The ratio of \textit{I$_{CTAT}$} to \textit{I$_{PTAT}$} results in a linear transfer characteristic (versus temperature)~\cite{Pertijs2005Feb}. The channel dual-slope MADC performs three tasks. First, it computes the ratio of \textit{I$_{CTAT}$} to \textit{I$_{PTAT}$}. Next, it multiplies this ratio by the PID coefficients, \textit{(c$_{0}$, c$_{1}$, c$_{2}$)}, and finally it subtracts the results from the scaled version of the desired temperature, \textit{(c$_{0}$\textit{t(k)}, c$_{1}$\textit{t(k)}, c$_{2}$\textit{t(k)})}. The result of these three operations is the computation of \textit{c$_{n}$e(k)}, as shown in~Fig. \ref{fig:VLSI_LOOP}. Next, these values are stored in the on-chip 7-bit SRAM bank. An on-chip adder adds the appropriate values according to Eq.(3) and computes \textit{u(k)}, which sets the duty cycle of the signal generated by the digital PWM. The duty cycle of this signal sets the amount of time that the in-cell heater is on, thus regulating the temperature. The in-cell heater was implemented by a PMOS transistor with a width and length of 100$\mu$m and 0.13$\mu$m, respectively, and a 15$\Omega$ load.

\begin{figure}[!ht]
	\centering
	\includegraphics[width=0.4\textwidth]{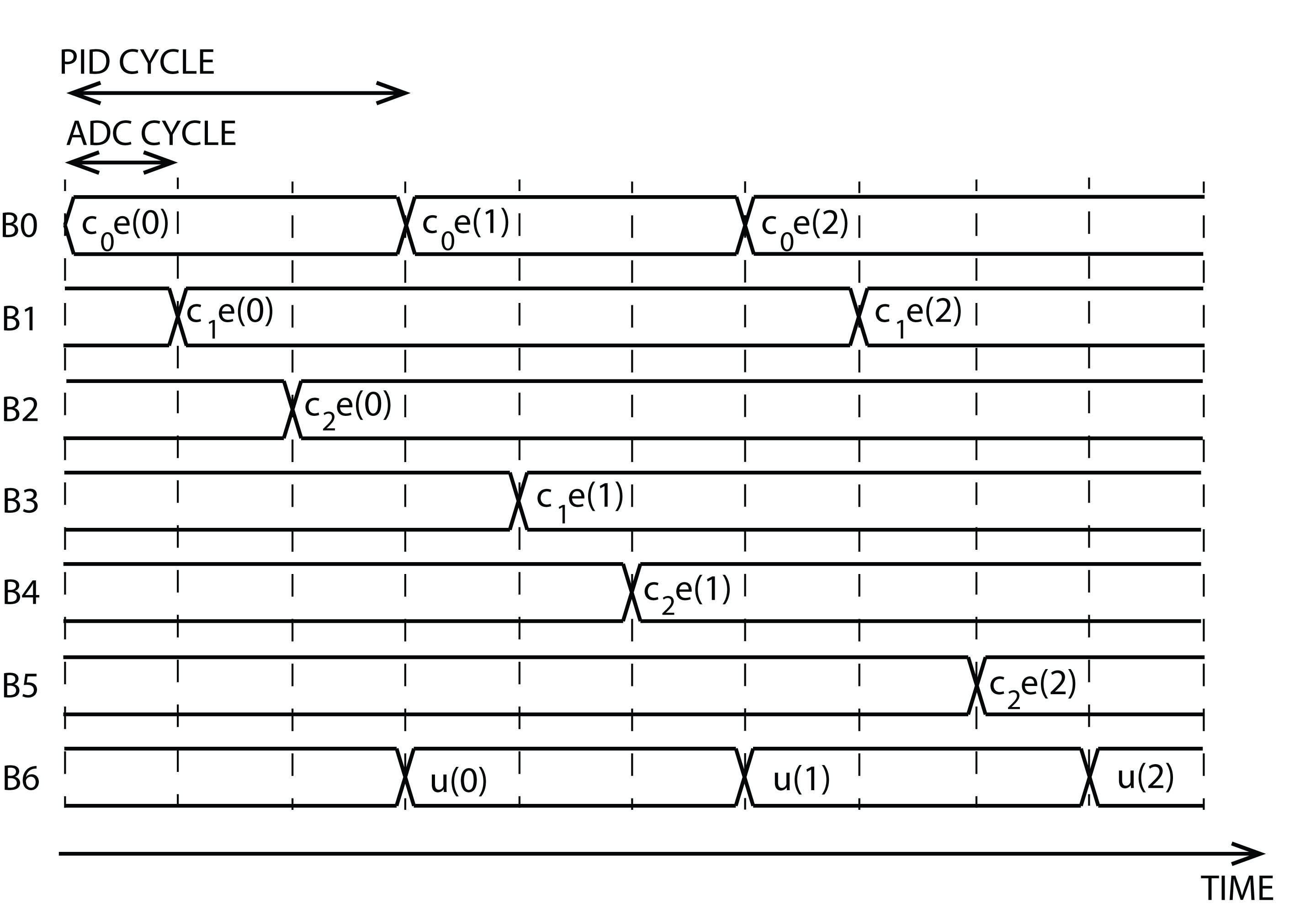}
	\caption{Timing diagram of the temperature regulation loop.}
	\label{fig:temp_PID_timing}
\end{figure}

The timing diagram of the PID regulation loop is shown in~Fig. \ref{fig:temp_PID_timing}. The channel dual-slope MADC performs three conversion cycles during one cycle of the PID controller, thus eliminating the need for three parallel ADCs. As shown in~Fig. \ref{fig:temp_PID_timing}, in the first conversion cycle of the PID controller, the ADC computes \textit{c$_{0}$e}(0), \textit{c$_{1}$e}(0), and \textit{c$_{2}$e}(0). Next, \textit{u}(0) is calculated from these values, and the duty cycle of the digital PWM is set accordingly. Consecutive values of \textit{c$_{0}$e(n)}, \textit{c$_{1}$e(n)}, \textit{c$_{2}e(n)$}, and \textit{u(n)} are generated in the same manner.

\section{Circuit Implementation}

\subsection{Current-to-Digital Conversion Channel}

\begin{figure}[!ht]
	\centering
	\includegraphics[width=0.5\textwidth]{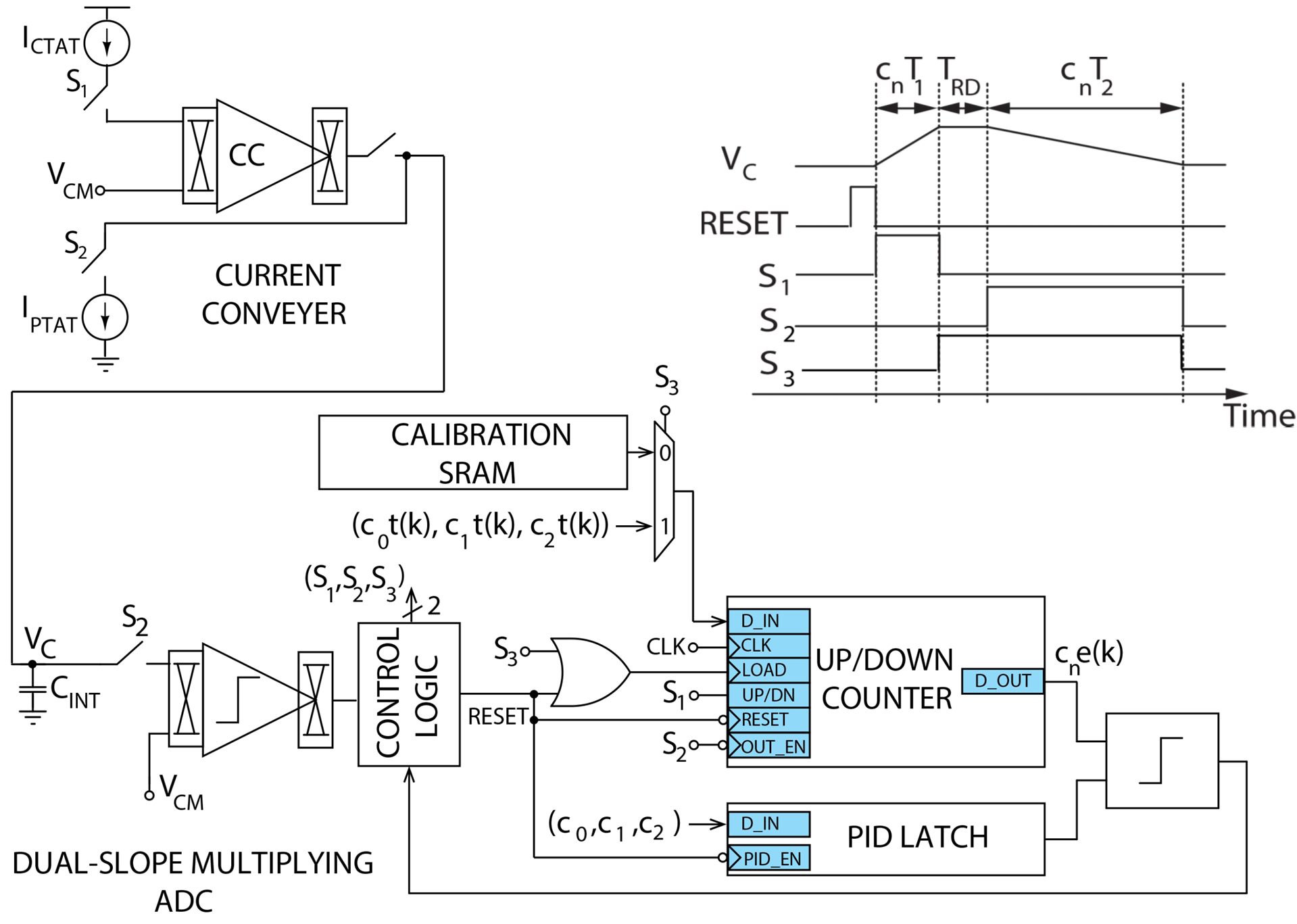}
	\caption{Top-level VLSI architecture of one temperature sensing channel and the timing diagram of the dual-slope MADC.}
	\label{fig:CHANNEL}
\end{figure}

The top-level VLSI architecture of the current-to-digital conversion channel is shown in~Fig.~\ref{fig:CHANNEL}.
Each channel consists of a chopper-stabilized current conveyor \cite{ISCAS}, a dual-slope MADC \cite{BIOCAS}, control logic, on-chip memories, and a counter. The dual-slope MADC performs both multiplication and subtraction, as required by the PID algorithm.

\begin{figure*}[!ht]
	\centering
	\includegraphics[width=1\textwidth]{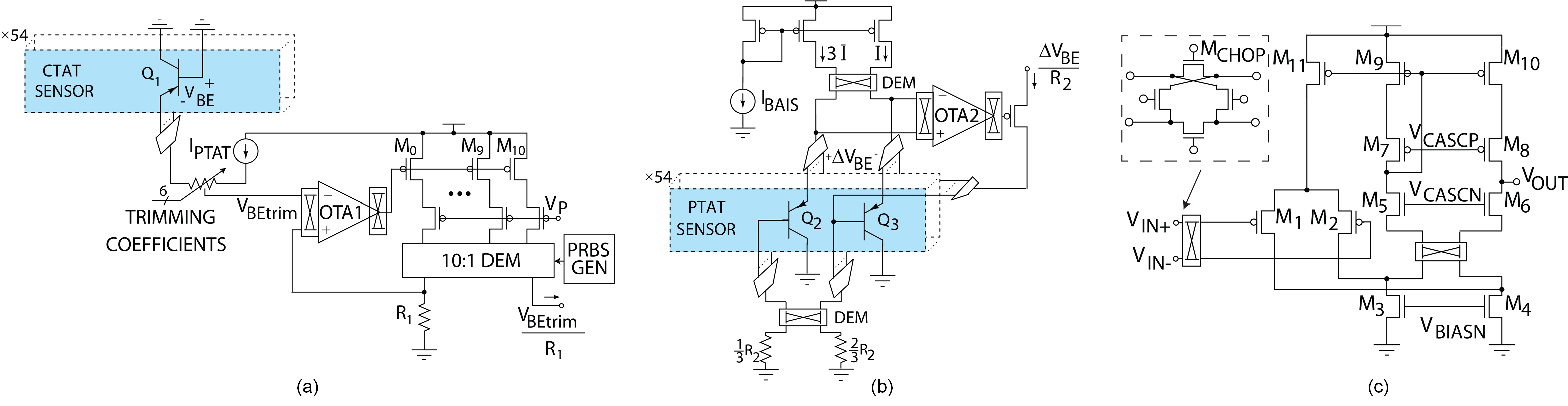}
	\caption{Circuit diagrams of (a) the CTAT sensor and current source, (b) the PTAT sensor and current source, and (c) the chopper-stabilized folded-cascode OTA in the CTAT (OTA1) and PTAT (OTA2) current sources.}
	\label{fig:ckt_CTAT_PTAT}
\end{figure*}

The timing diagram of the MADC for a typical conversion cycle is shown in Fig. \ref{fig:CHANNEL}. The dual-slope MADC digitizes the input signal in two phases: charging phase \textit{T$_{1}$} and discharging phase \textit{T$_{2}$}. The ratio of the duration of the charging phase to the discharging phase represents the input signal level. As shown in Fig. \ref{fig:CHANNEL}, to implement multiplication of the input current by a digital PID coefficient \textit{c$_{n}$}, the duration of the charging phase is scaled with a constant coefficient \textit{c$_{n}$}$<$1~\cite{BIOCAS}. In this case, by counting the time \textit{c$_{n}$}\textit{T$_{2}$}, a digital representation of \textit{c$_{n}$}\textit{I$_{CTAT}$} can be obtained.

As shown in Fig.~\ref{fig:CHANNEL}, the counter is first reset. At the same time, the PID multiplication coefficient, \textit{c$_{n}$}, is loaded into the in-channel latch (\textit{S$_{3}$=1}). There is a large channel-to-channel variation between the BJT current output. To compensate for this variation, the channel counter is pre-loaded with a digital calibration value (stored in the in-channel calibration SRAM) at the rising edge of the RESET signal. Next, the in-channel counter counts up from this value to time \textit{c$_{n}$T$_{1}$}, and the input current \textit{I$_{CTAT}$} is integrated onto the capacitor \textit{C$_{INT}$}. After time \textit{c$_{n}$T$_{1}$}, the voltage on the capacitor is held constant for a fixed time interval \textit{T$_{RD}$}. During this time, the digital representation of the desired temperature that is scaled by the PID coefficients \textit{c$_{n}$t(k)} is loaded into the counter. During time \textit{c$_{n}$T$_{2}$}, the integrating capacitor is discharged with the PTAT current source. During time \textit{c$_{n}$T$_{2}$}, the counter counts down, and the final value of the counter is available at the falling edge of \textit{S$_{2}$} signal. By counting down from \textit{c$_{n}$t(k)} in the second phase of the conversion cycle, the MADC performs subtraction of \textit{c$_{n}$y(k)} from \textit{c$_{n}$t(k)} (as shown in Fig.~\ref{fig:VLSI_LOOP}) thus computing the error signal \textit{c$_{n}$e(k)}.

The front-end bidirectional current conveyor is implemented by a PMOS and an NMOS transistor connected in the feedback of an OTA \cite{TCAS}. The negative feedback ensures a known potential at the working electrode is set by the voltage at the non-inverting input of the OTA. It also enables the current conveyor to source and sink an input current without the need for a DC offset current. The OTA is implemented as a folded-cascode amplifier with a PMOS input pair. Chopper-stabilization is utilized to reduce the OTA flicker noise and offset and dynamic element matching is utilized to reduce the output current mirrors mismatch \cite{TCAS}.

The ADC comparator is implemented with three stages of pre-amplifiers with a total gain of 60dB, and a high-speed latch as the last stage. The first stage of the comparator is implemented as a cross-coupled diode-connected gain stage to provide a moderate gain with high bandwidth. Chopper-stabilization suppresses the input offset and ensures a 9-bit accuracy. The second and third stages are identical to the first one but with no chopping. The high-speed latch is implemented with an NMOS input pair gain stage and an NMOS-PMOS cross-coupled load to provide high accuracy, low offset and a high bandwidth \cite{TBIOCAS}.

\subsection{CTAT and PTAT Current Sources}

A simplified circuit diagram of the CTAT current source is shown in Fig.~\ref{fig:ckt_CTAT_PTAT}(a). To generate a digitally programmable base-emitter voltage \textit{V$_{BEtrim}$} (compensating for chip-to-chip variations), a PTAT current (generated by a separate on-chip biasing circuit) is passed through a 6-bit programmable resistor connected in series with the diode-connected pnp transistor \textit{Q$_{1}$}. The current \textit{V$_{BEtrim}/$R$_{1}$} is generated by a voltage-to-current converter, as shown in Fig.~\ref{fig:ckt_CTAT_PTAT}(a). A large resistor (\textit{R$_{1}$}$>$100M$\Omega$) is required to ensure that the CTAT current is kept relatively low (20nA) and stays within the dynamic range of the ADC. This would result in a large integration area. To reduce the size of the resistor, a current mirror with a 10:1 ratio is utilized. A 10:1 dynamic element matching is used to reduce the effect of the current mirror mismatch. In this work \textit{R$_{1}$} is set to 1.5M$\Omega$.

A simplified circuit diagram of the PTAT current source is shown in~Fig. \ref{fig:ckt_CTAT_PTAT}(b). The \textit{$\Delta$V$_{BE}$} is generated by determining the difference between the \textit{V$_{BE}$} of the two substrate pnp transistors that are biased at 3:1 collector current ratio. The voltage-to-current converter consisting of OTA2 and the NMOS feedback transistor generates \textit{$\Delta$V$_{BE}$} across \textit{R$_{2}$} (split into \textit{R$_{2}$$/$3} and \textit{2R$_{2}$$/$3}). In this work \textit{R$_{2}$} is set to 100k$\Omega$. This results in the desired output current \textit{$\Delta$V$_{BE}/$R$_{2}$}. The resistor \textit{R$_{2}/$3} is added in series with the base of the \textit{Q$_{3}$} transistor in order to ensure that the base current of the \textit{Q$_{3}$} does not affect the output current.

The main causes of inaccuracy in the current sources are the input offset and the flicker noise of OTAs and the mismatch of the bias current mirror. Internal OTA chopping is utilized to reduce the effect of flicker noise and input offset voltage \cite{Sensors2021}. OTA1 and OTA2 adopt the same circuit topology, as shown in Fig. \ref{fig:ckt_CTAT_PTAT}(c). The transistor sizes are given in~Table~\ref{OTASIZE}. 
The chopper switches are placed at the input of the OTA and below the cascoded NMOS transistors in the current mirror. This placement reduces the flicker noise and DC offsets caused by the input pair transistors and the NMOS current mirror transistors. Dynamic element matching at 100Hz, by the means of the chopper switches in series between the current sources and the resistors, is utilized to improve the matching. 

\begin{table}[!ht]
	\renewcommand{\arraystretch}{1}
	\centering
	{\small
	\caption{OTA Transistor Sizing}
	\label{OTASIZE}
	\begin{tabular}{ccccc}
		\hline
    	Transistor  &	     W/L $(\mu m)$\\
       \hline
		$M_{1,2}$   &	$9\times 4 / 1$  \\
		$M_{3,4}$   &	$4\times 1.5 / 4 $ \\
		$M_{5,6}$   &	$5\times 1.5 / 1 $ \\
		$M_{7,8}$   &	$10\times 1.5 / 1 $\\
		$M_{9,10}$  &	$4\times 1.5 / 4 $ \\
		$M_{11}$    &	$11\times 1.5 / 2 $  \\
		\bottomrule
	\end{tabular}
	}
\end{table}

\begin{figure}[!ht]
\centering
    \includegraphics[width=0.35\textwidth]{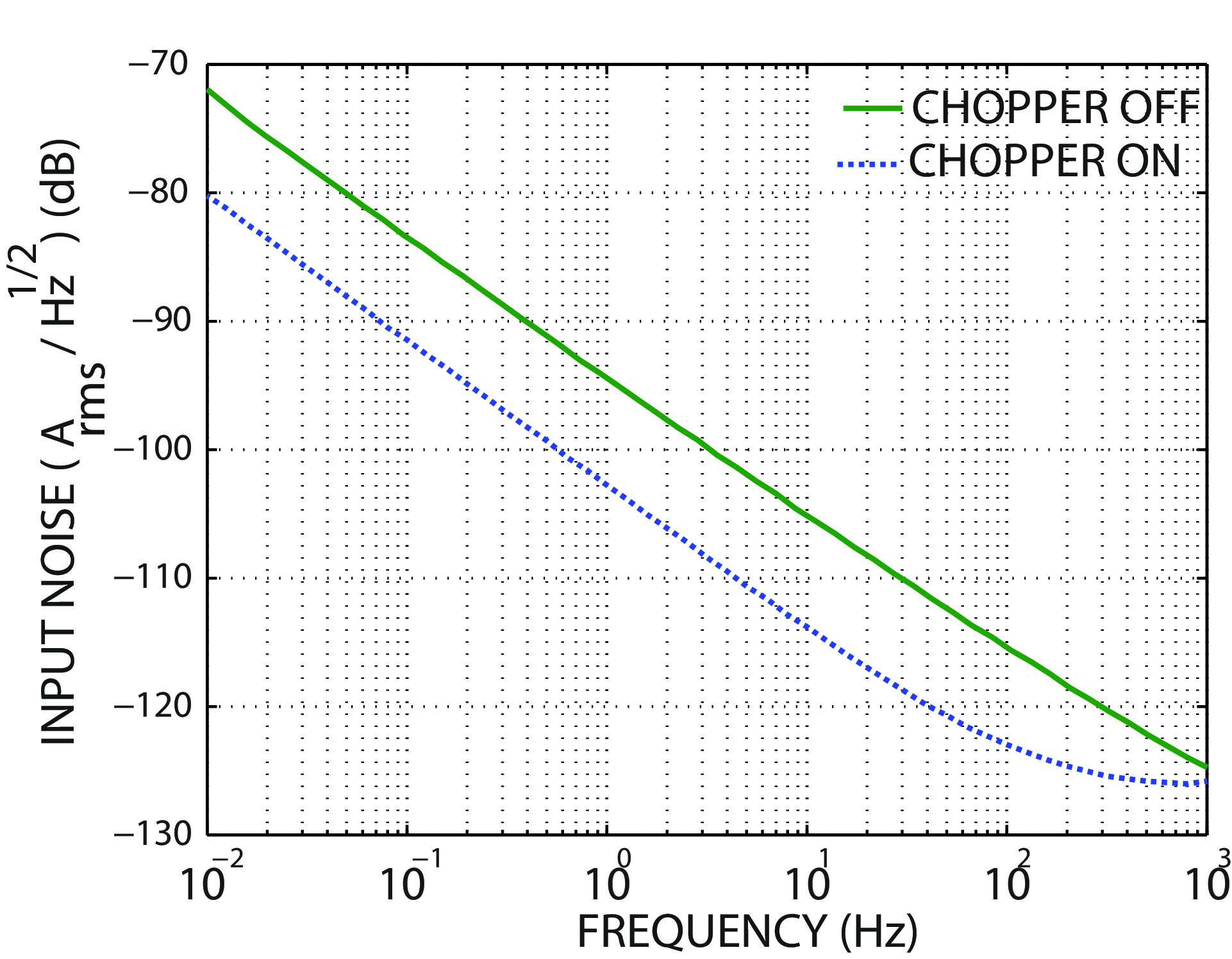}
	\caption{Simulated input-referred noise spectrum of the OTA with chopper enabled and disabled from 0.01Hz to 1kHz.}
	\label{fig:OTANOISE}
\end{figure}

\begin{figure}[!ht]
   \centering
   \leavevmode
   \epsfig{file=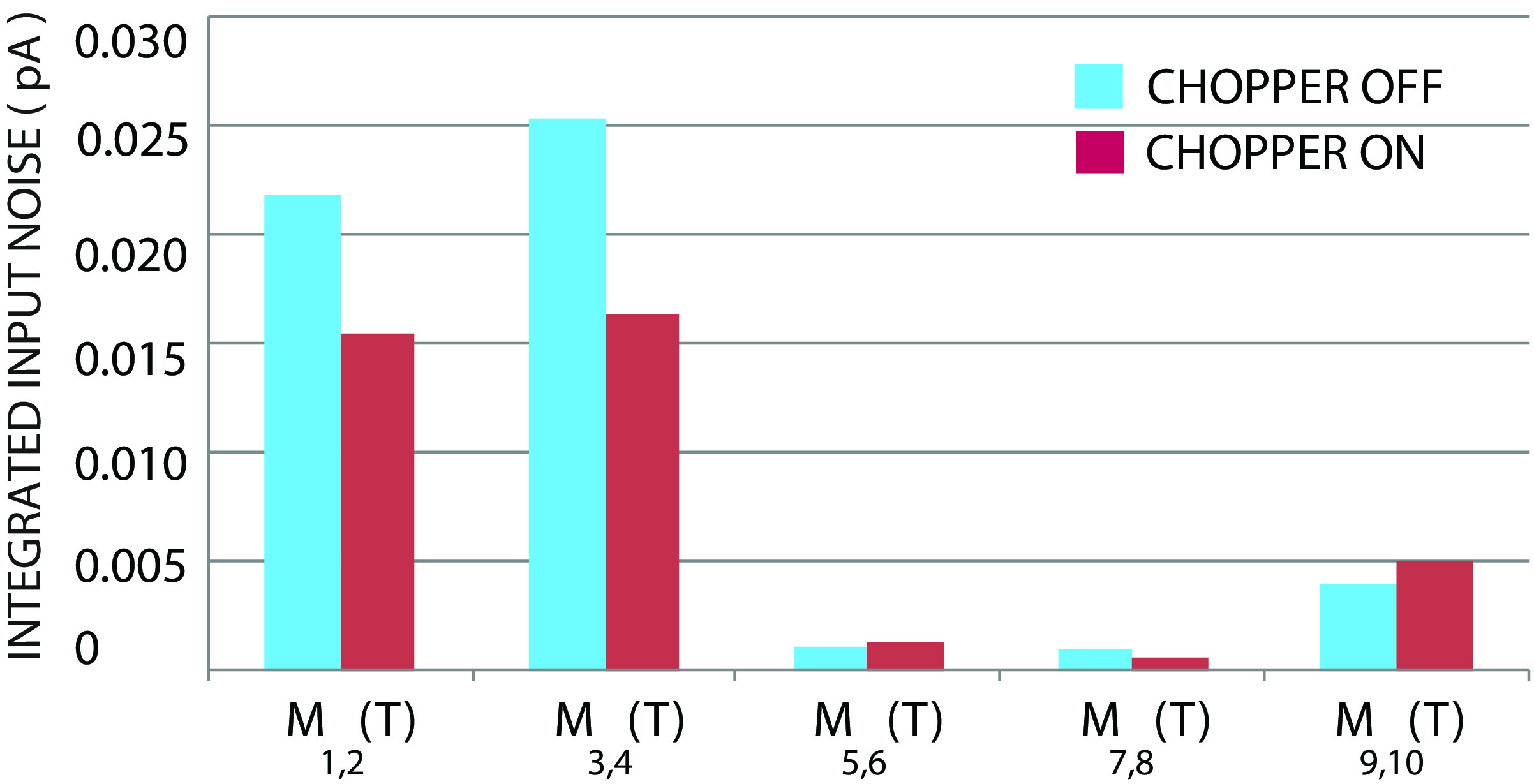,width= 0.35\textwidth}\\
{\it (a)}\\
   \vspace{0.1in}
   \epsfig{file=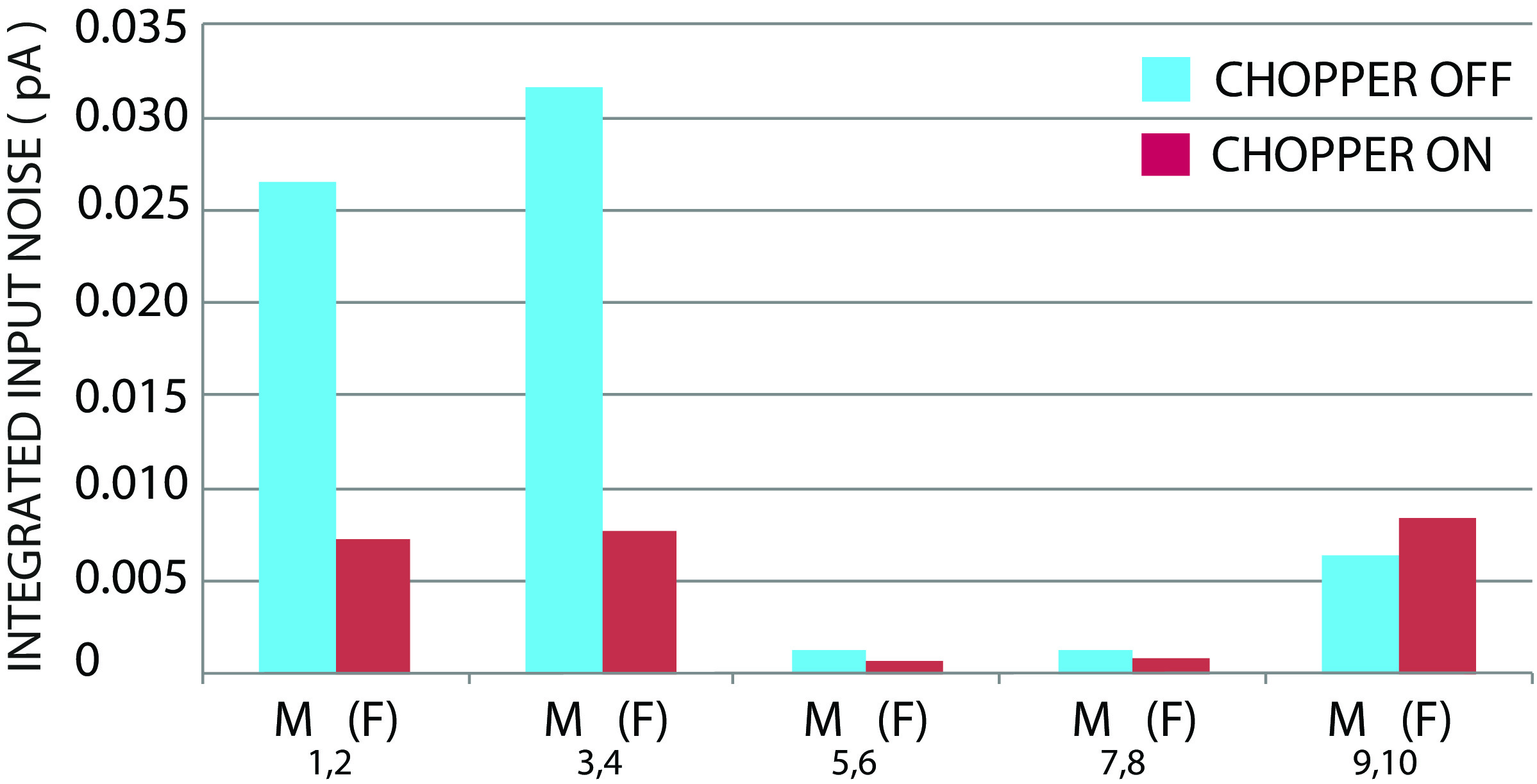,width=0.35\textwidth}\\
{\it (b)}
   \caption{Simulated OTA noise. (a) Thermal noise contribution, and (b) flicker noise contribution.}
   \label{fig:NoiseTab}
\end{figure}

The simulated input-referred noise of OTA1 in cases where the chopper is disabled and enabled is shown in~Fig.~\ref{fig:OTANOISE}. 
The integrated input-referred noise from 0.01Hz to 1kHz is 0.11pA when the chopper is disabled and 0.06pA when the chopper is enabled. The contribution of each transistor to the total input-referred noise is shown in~Fig.~\ref{fig:NoiseTab}. When the chopper is disabled, the main contributions to the input-referred flicker noise are from the OTA current mirror transistors \textit{M$_{3}$} and \textit{M$_{4}$}, and from the input pair transistors \textit{M$_{1}$} and \textit{M$_{2}$}. When the chopper is enabled, the current mirror transistors \textit{M$_{9}$} and \textit{M$_{10}$} are the main contributors to the input-referred flicker noise.

\subsection{Digital Pulse Width Modulator}

\begin{figure}[!ht]
	\centering
	\includegraphics[width=7cm]{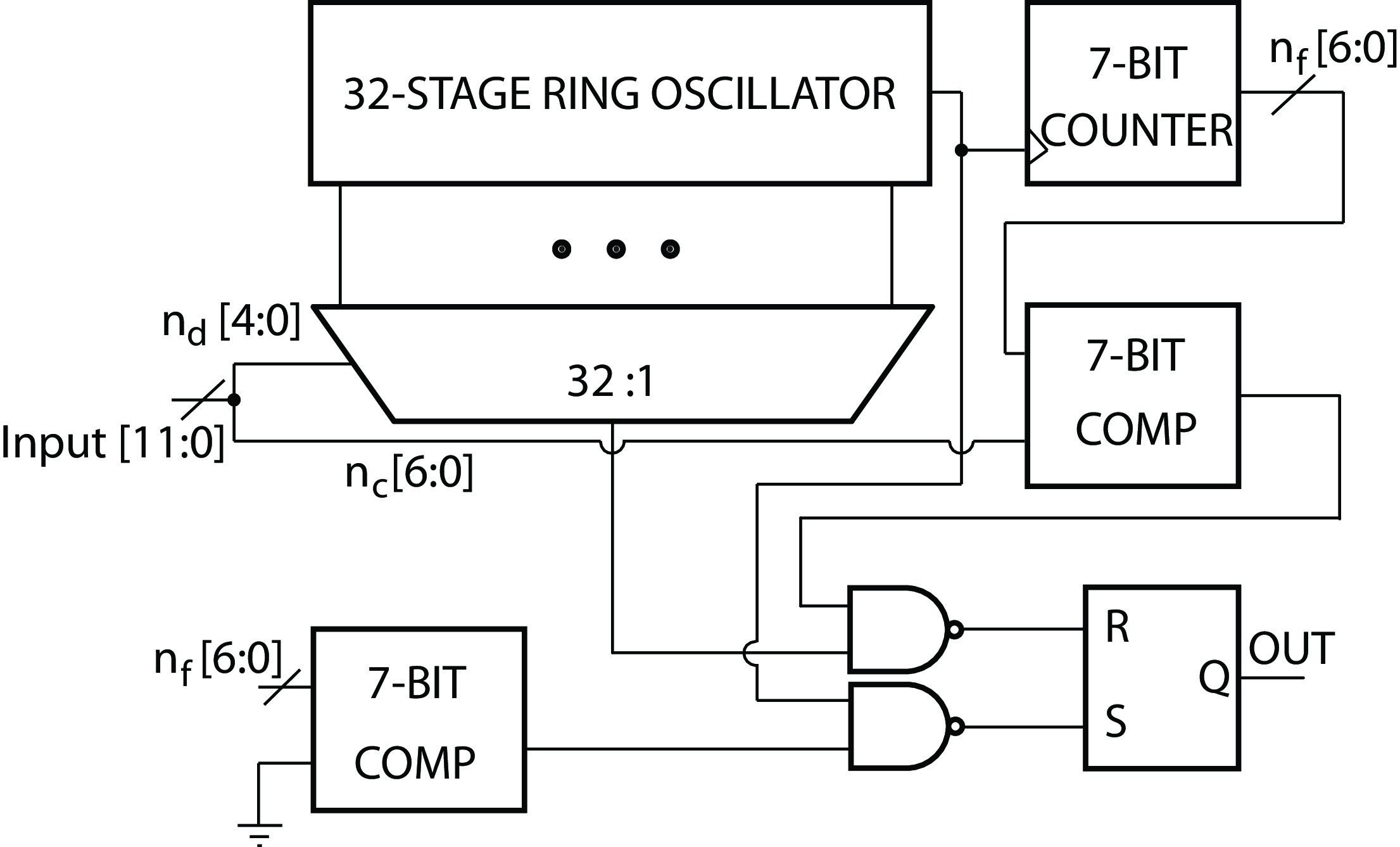}
	\caption{Block diagram of the digital pulse width modulator.}
	\label{fig:DPWM}
\end{figure}

In the temperature regulation loop, a digital PWM sets the duty cycle of the pulse for controlling the in-channel heater. The achievable discrete set of the duty cycle settings of the pulse depends on the digital PWM resolution. In this work, a 12-bit digital PWM architecture is selected. The design is based on a hybrid delay-line/counter architecture in~\cite{PP}. The block diagram of the PWM is shown in~Fig. \ref{fig:DPWM}. In this architecture, a 7-bit counter and 32-stage ring oscillator are used to achieve the 12-bit resolution. At the beginning of a switching cycle, the output set-reset (SR) flip-flop is set, and the PWM output pulse goes high. The pulse that propagates through the ring at the oscillation frequency serves as the clock for the counter. At the time when the counter output matches the top most significant bits of the digital input (\textit{n$_{c}$}), and a pulse reaches the tap selected by the least significant bits (\textit{n$_{d}$}), the output flip-flop is reset, and the output pulse goes low~\cite{PP}.

\section{Experimental Results}

\subsection{Chip Micrograph}

\begin{figure}[!ht]
	\centering
	\includegraphics[width=7.5cm]{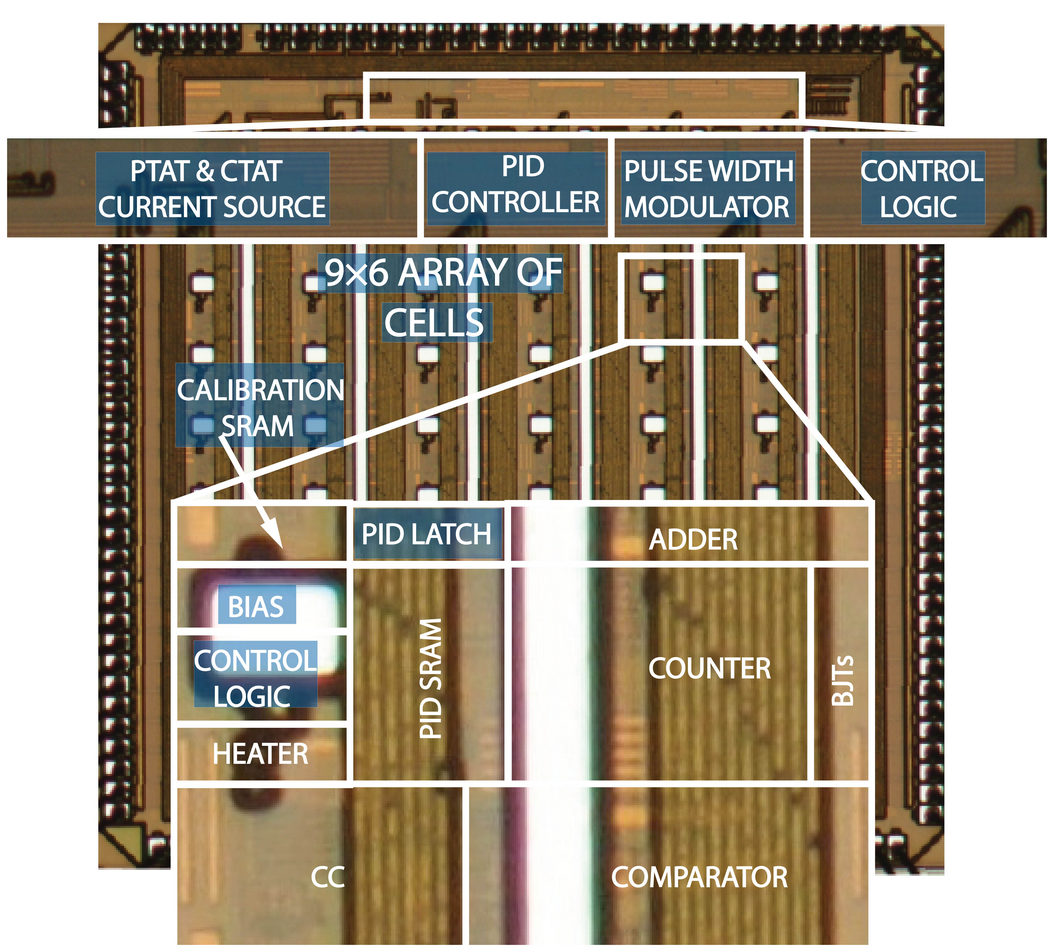}
	\caption{Micrograph of the wireless temperature-regulated DNA analysis SoC fabricated in 0.13$\mu$m CMOS technology.}
	\label{fig:DIE}
\end{figure}

The micrograph of the fabricated multi-modal amperometric DNA analysis SoC is shown in Fig.~\ref{fig:DIE}.
The 54 cells are arranged in a 9$\times$6 array on a 3mm$\times$3mm CMOS die. Each cell consists of a current conveyor, a dual-slope MADC, an in-cell bias and clock generator, a pH sensor, a heater, and temperature sensing BJTs. The PID controller, CTAT and PTAT current sources, and the digital PWM are located in the top section of the die. The electrodes are postprocessed with 2D and 3D Au top layers and several metal underlayers \cite{JSSC}.


\subsection{Temperature Regulation Testing Results}

\begin{figure}[!ht]
	\centering
	\includegraphics[width=0.35\textwidth]{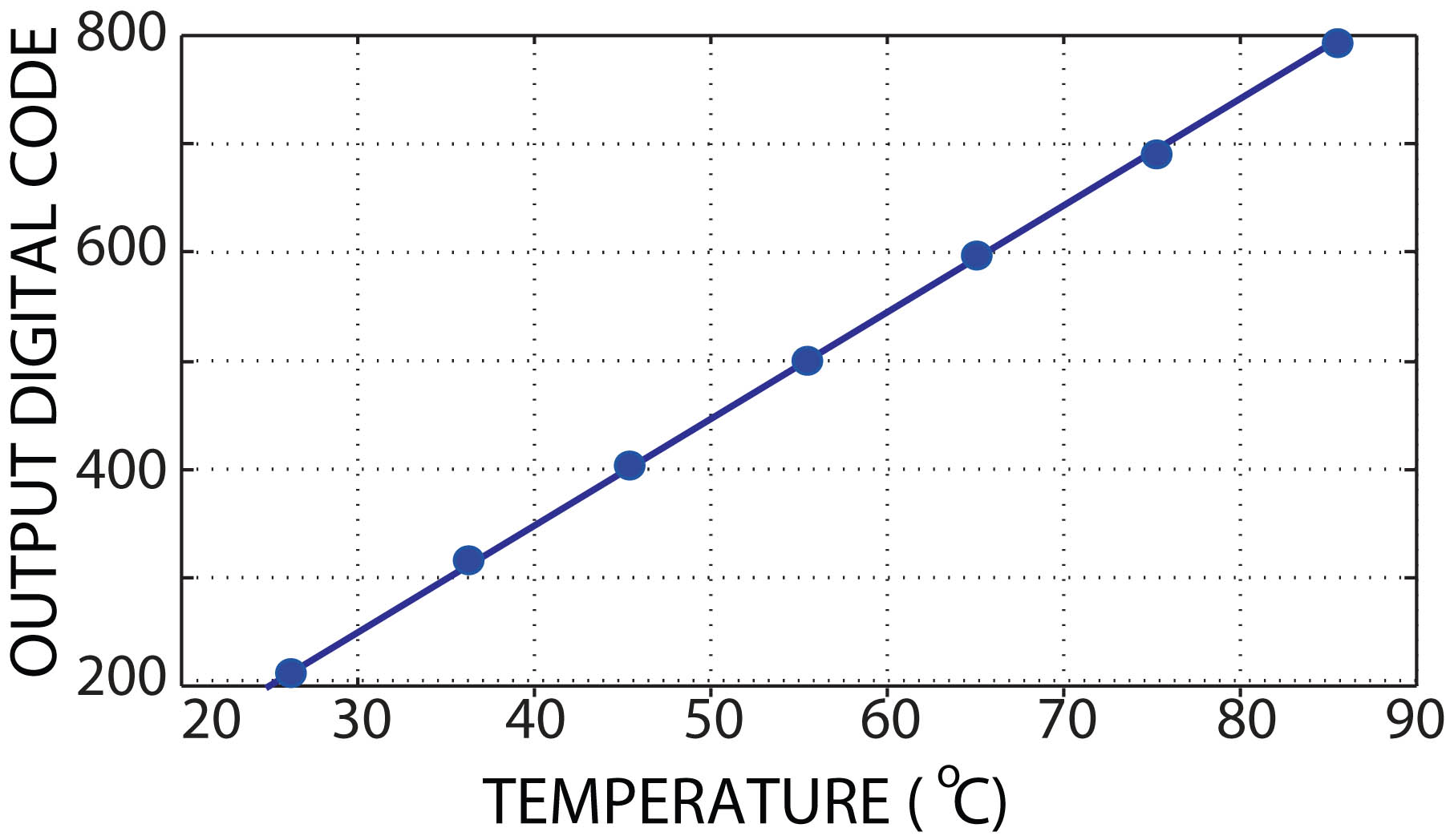}
	\caption{Experimentally measured digital output of the ADC v.s. temperature.}
	\label{fig:TEMP_LIN}
\end{figure}

\begin{figure}[!ht]
	\centering
	\includegraphics[width=0.35\textwidth]{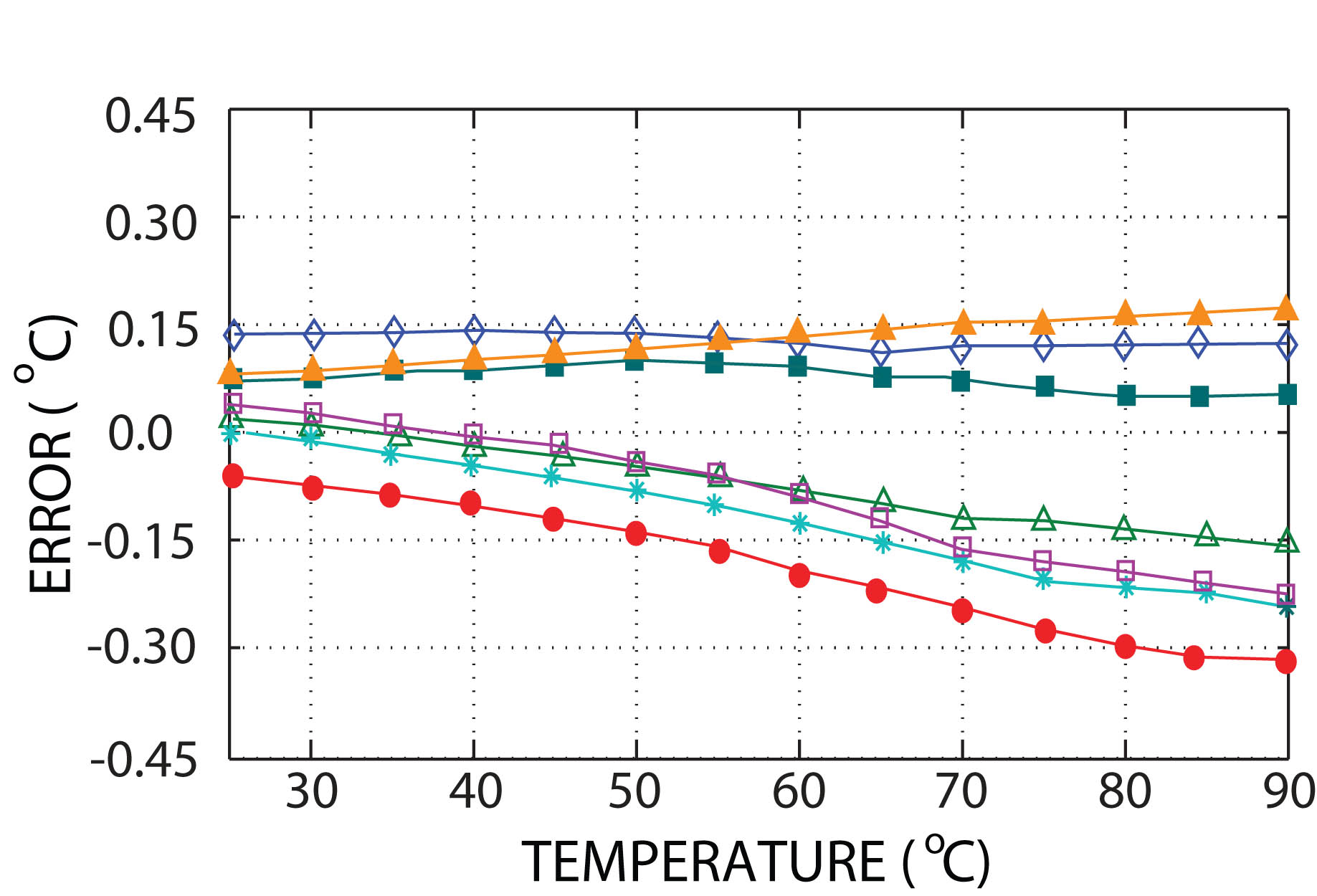}
	\caption{Temperature regulation error experimentally measured on seven dies from one wafer.}
	\label{fig:TEMP_ERR}
\end{figure}

The output digital code of the ADC versus temperature, from 20$^{\circ}$C to 90$^{\circ}$C, is shown in~Fig.~\ref{fig:TEMP_LIN}. 
The temperature sensor archives a linear transfer function over the operating temperature range. Fig.~\ref{fig:TEMP_ERR} shows the measured temperature error of seven dies from one wafer, which operated at a supply voltage of 1.2V. After a one-point calibration for compensating for the BJT variation across channels, the temperature error across the seven dies was less than ±0.5$^{\circ}$C within the temperature range of 20$^{\circ}$C to 90$^{\circ}$C. 

\begin{figure}[!ht]
	\centering
	\includegraphics[width=0.35\textwidth]{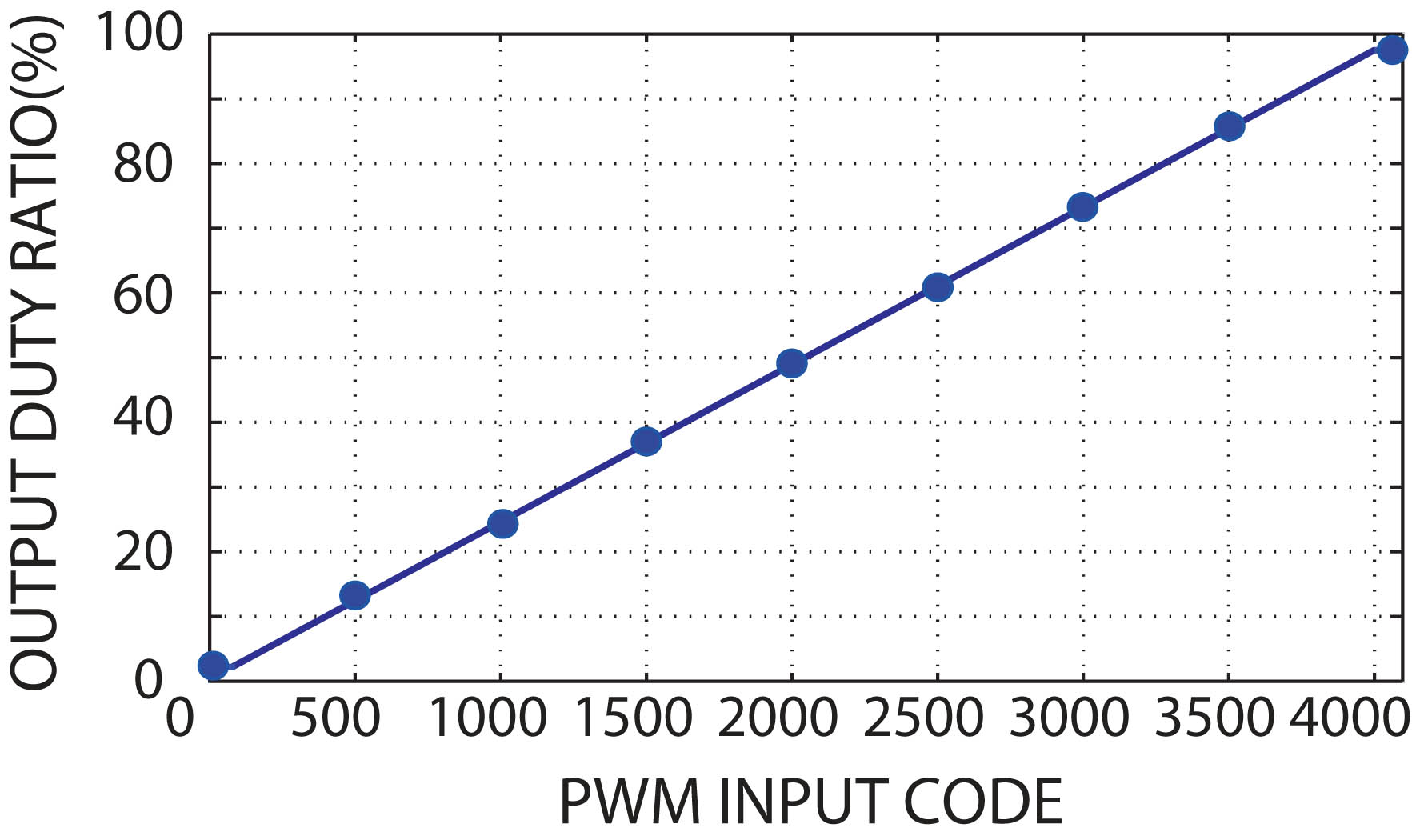}
	\caption{Experimentally measured transfer function of the pulse-width-modulator.}
	\label{fig:PWMD}
\end{figure}

The digital PWM is characterized by sweeping the digital input code from 0 to 4095 and clocking the digital PWM at 2.5MHz. The experimental results of~Fig.~\ref{fig:PWMD} show the measured duty ratio of the output pulses as a function of the 12-bits digital input. The minimum (4$\%$) and the maximum (96$\%$) duty ratios are set by design. The digital PWM achieves a linear transfer function over the programmable digital input code with a 0.82 percent maximum error. The digital PWM achieves a minimum duty cycle resolution of 0.1$\mu$S.

\begin{figure}[!ht]
	\centering
	\includegraphics[width=0.35\textwidth]{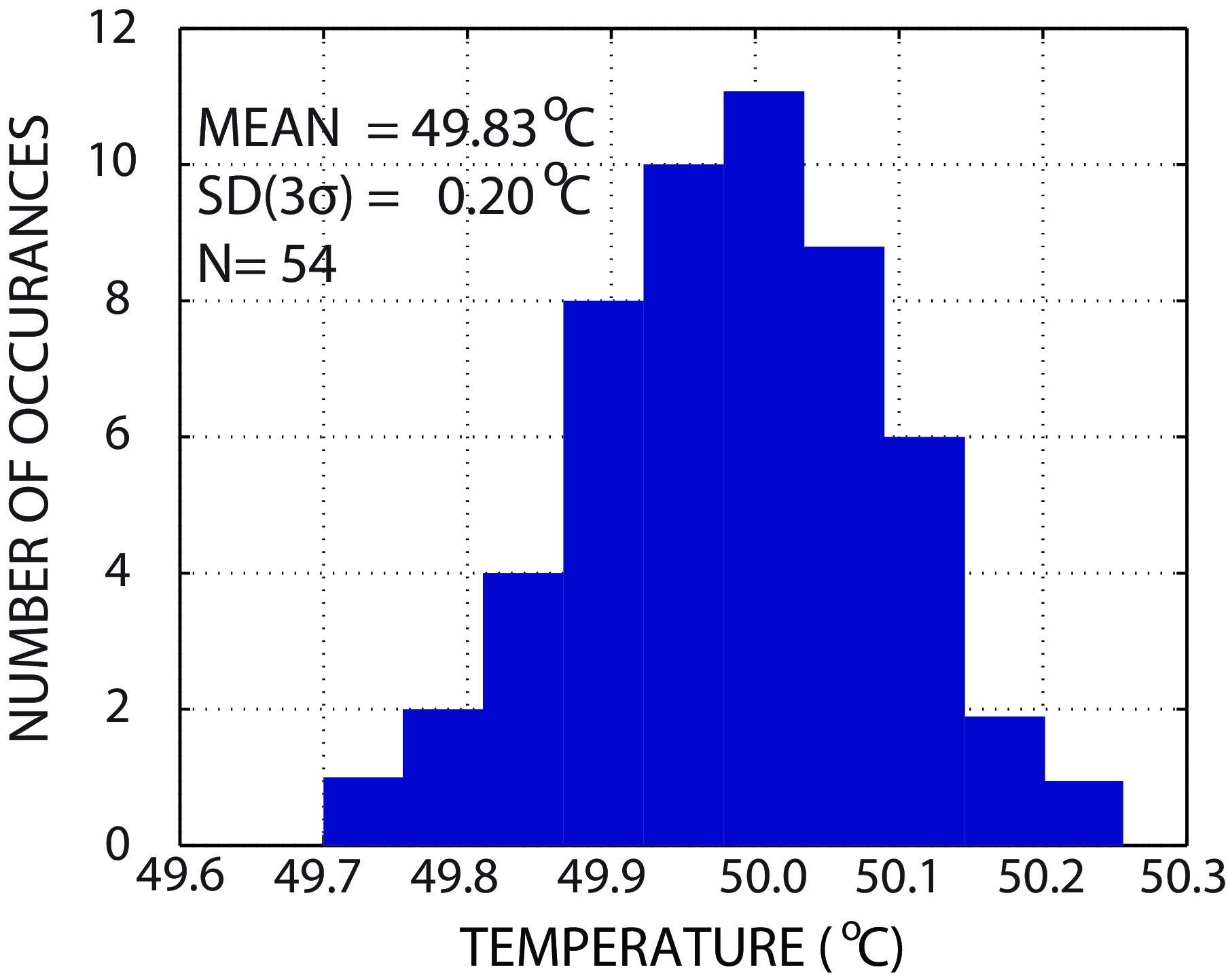}
	\caption{Experimentally measured temperature from 54 channels in one chip.}
	\label{fig:TEMP_ERR2}
\end{figure}

An external heater was utilized to set the chip temperature at 50$^{\circ}$C and the temperature at each channel was recorded to study the effect of channel-to-channel mismatch. Experimentally measured temperature from 54 channels in one chip is shown in Fig.~\ref{fig:TEMP_ERR2}. The mean digital output temperature and the corresponding standard deviation are 49.83$^{\circ}$C and 0.20$^{\circ}$C, respectively.

\begin{figure}[!ht]
	\centering
	\includegraphics[width=0.35\textwidth]{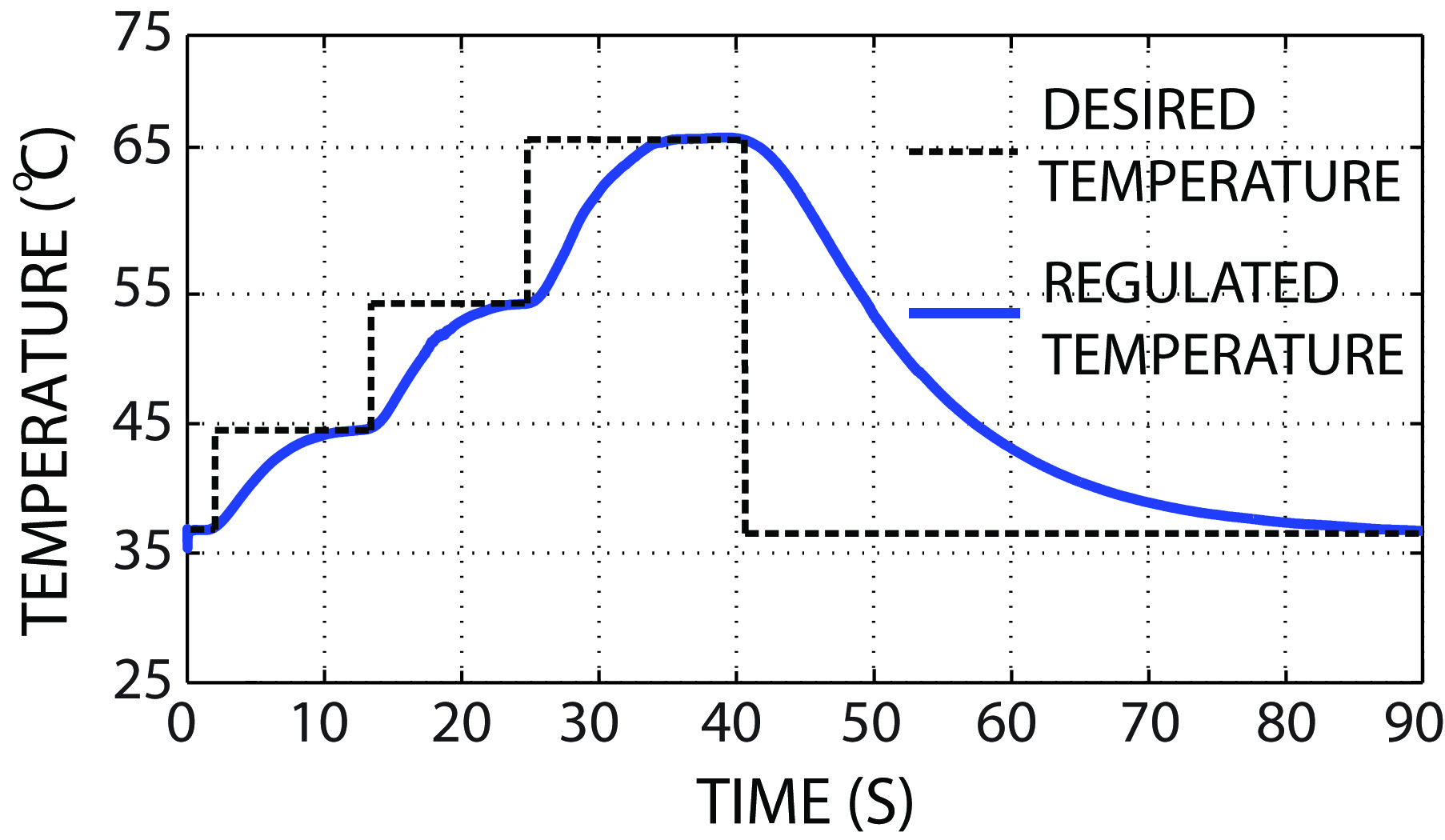}
	\caption{Transient responses of temperature regulation with target temperatures at 35$^{\circ}$C, 45$^{\circ}$C, 55$^{\circ}$C, and 65$^{\circ}$C.}
	\label{fig:TEM_REG}
\end{figure}

\begin{figure}
	\centering
	\includegraphics[width=0.35\textwidth]{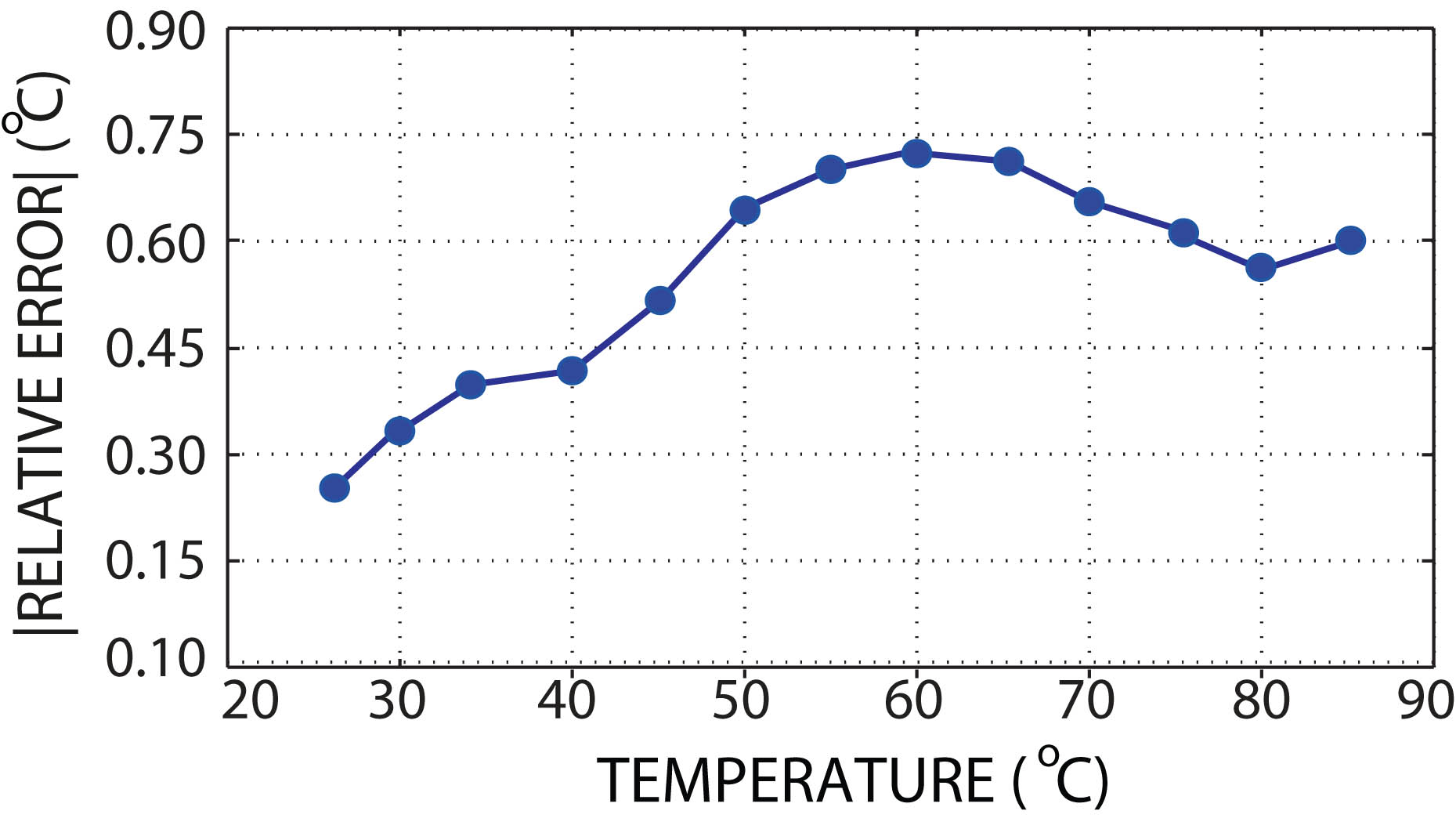}
	\caption{Measured absolute value of the relative error of the PID regulation loop.}
	\label{fig:TEMP_REG_ERR}
\end{figure}

An example of the temperature regulation cycle in liquid (5mL 1M potassium phosphate buffer solution), with steps at 35$^{\circ}$C, 45$^{\circ}$C, 55$^{\circ}$C, and 65$^{\circ}$C, is shown in Fig.~\ref{fig:TEM_REG}. The solid line is the chip temperature and the dashed line is the desired temperature. It takes roughly 10 seconds to achieve an increase of 5$^{\circ}$C in the chip. The measured absolute value of the relative error of the PID regulation loop over the operating temperature range is shown in Fig.~\ref{fig:TEMP_REG_ERR}. The absolute value of the error stays below 0.75$^{\circ}$C over the operating temperature range.

Table \ref{summaryTab2} provides a summary of the experimentally measured characteristics of the system for temperature regulation. 
\begin{table}[!ht]
	\centering
	\renewcommand{\arraystretch}{1.1}
	{\footnotesize
	\caption{Experimentally Measured Characteristics}
	\label{summaryTab2}
	\begin{tabular}{lcc}
		\hline
		Technology		                &  0.13$\mu$m CMOS \\
		Supply Voltage	               	&  1.2V \\
		Area	     	                &  3mm$\times$3mm \\
		Array Dimensions	            &  9$\times$6 channels \\
		Channel Size		            &  300$\mu$m$\times$200$\mu$m \\
		Sensitivity		                &  8.6pA \\
        Power Consumption (System)	&      \\
		\hspace{3 mm}SRAM            	&  1.3$\mu$W \\
	    \hspace{3 mm}PID Controller		&  21$\mu$W  \\
	    \hspace{3 mm}Temperature Core	&  12$\mu$W  \\
	    \hspace{3 mm}Heater	&  270mW @ 90$^{\circ}$C  \\
                      Power Consumption (Channel)	&      \\
     	\hspace{3 mm}Current conveyor  	&  8$\mu$W   \\
		\hspace{3 mm}Comparator       	&  19$\mu$W  \\
		\hspace{3 mm}Biasing          	&  4$\mu$W  \\
		\hspace{3 mm}Digital			&  11$\mu$W  \\
		\hspace{3 mm}Total (Channel)  	&  42$\mu$W \\	        	
		\hline
	\end{tabular}
	}
\end{table}

\subsection{Multi-modal Amperometric DNA Analysis Results}

The dynamic performance of the cell was measured with a 15Hz sinusoidal input current at a full-scale current of 400nA. The ADC was clocked at 10MHz. 
A SNR of 56.9dB was measured, leading to an effective number of bits (ENOB) of 9.3. The distortion was limited by a -58.7dB HD2 due to the single-ended architecture of the ADC. 

\begin{figure}[!ht]
	\centering
	\includegraphics[width=0.35\textwidth]{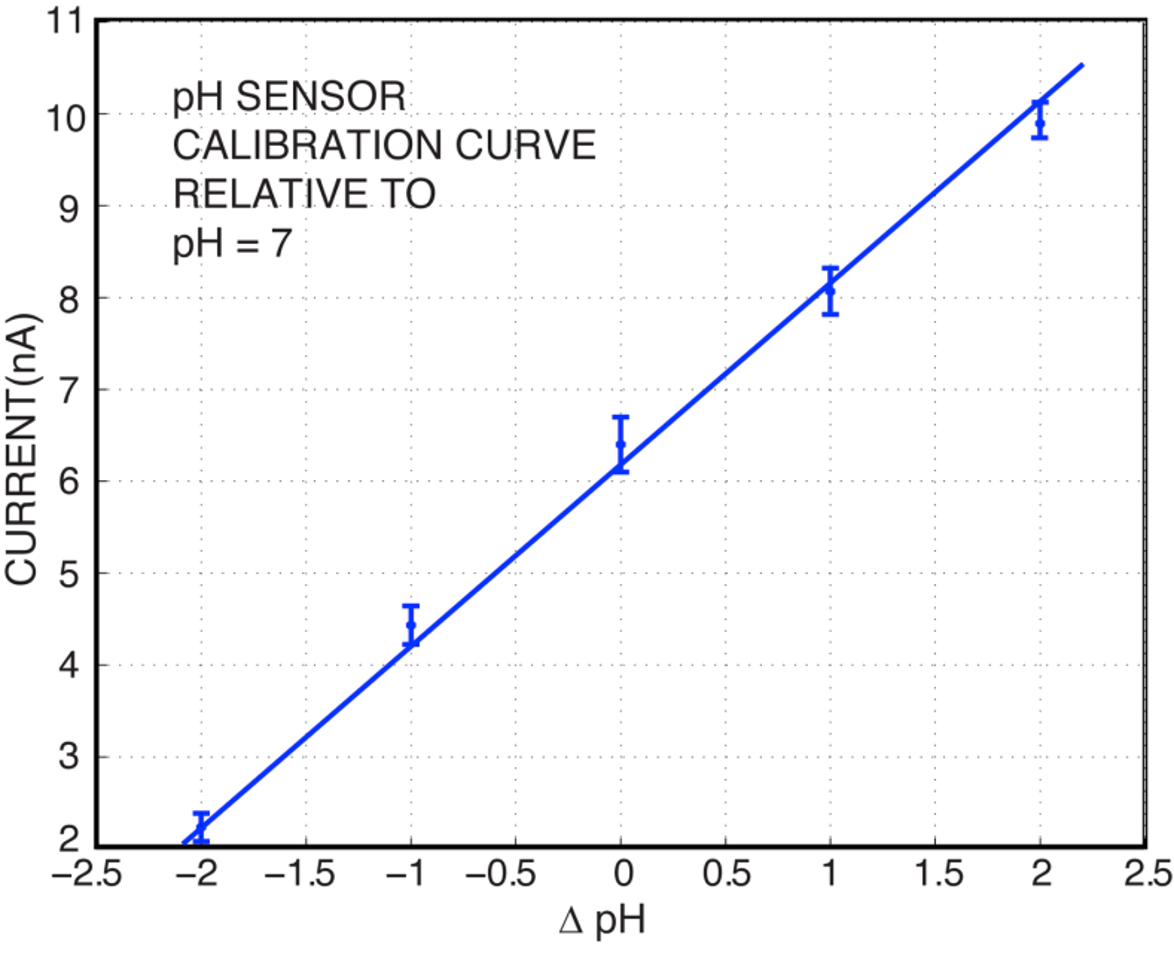}
	\caption{Experimental characterization of pH sensing in the CPA mode. The 3$\sigma$ error bars from 20 measurements are shown. }
	\label{fig:pH}
\end{figure}
Fig. \ref{fig:pH} shows the experimental characterization of pH sensing in the CPA mode. The 3$\sigma$ error bars from 20 measurements are shown in the figure. The measured current leads to a linear response of 1.8nA/pH within ±2 relative to a pH of 7.

\begin{figure}[!ht]
	\centering
	\includegraphics[width=0.35\textwidth]{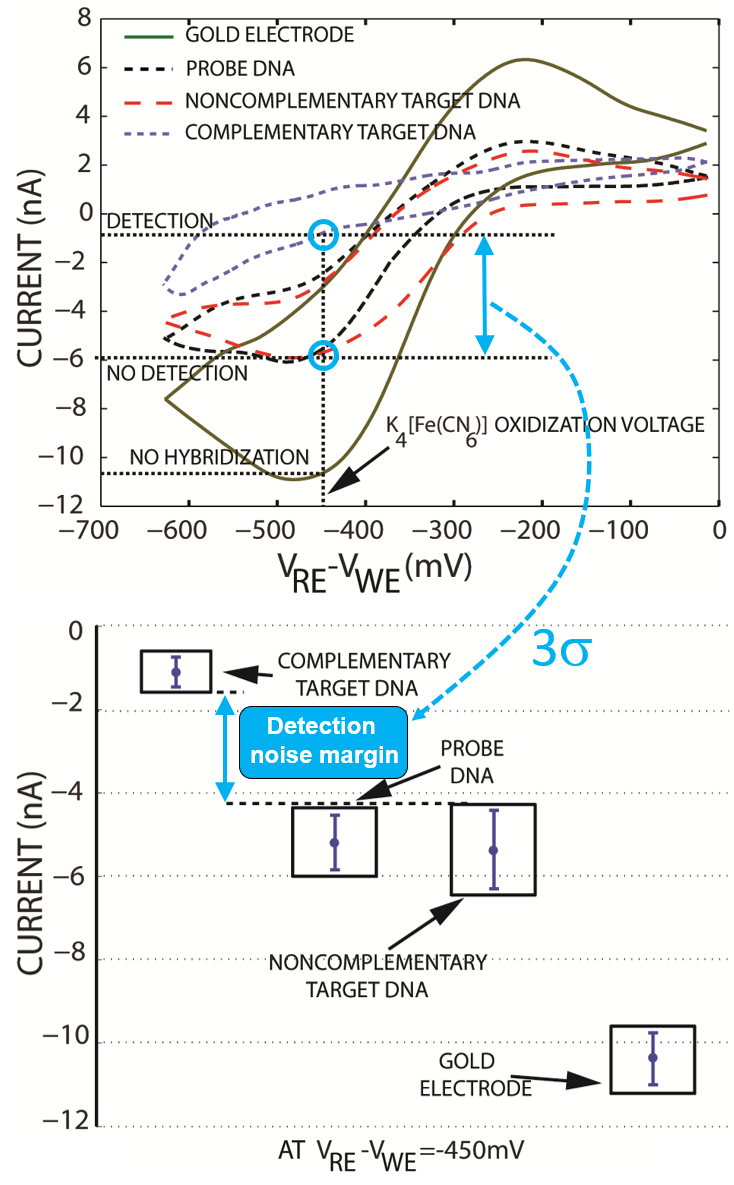}
	\caption{Experimental measurements of 5µM prostate cancer DNA using CV from 0 to -700mV at a scan rate of 100mV/s. The 3$\sigma$ error bars show 20 measurements from 3 chips.}
	\label{fig:exp_DNA_CV}
\end{figure}

\begin{figure}[!ht]
	\centering
	\includegraphics[width=0.35\textwidth]{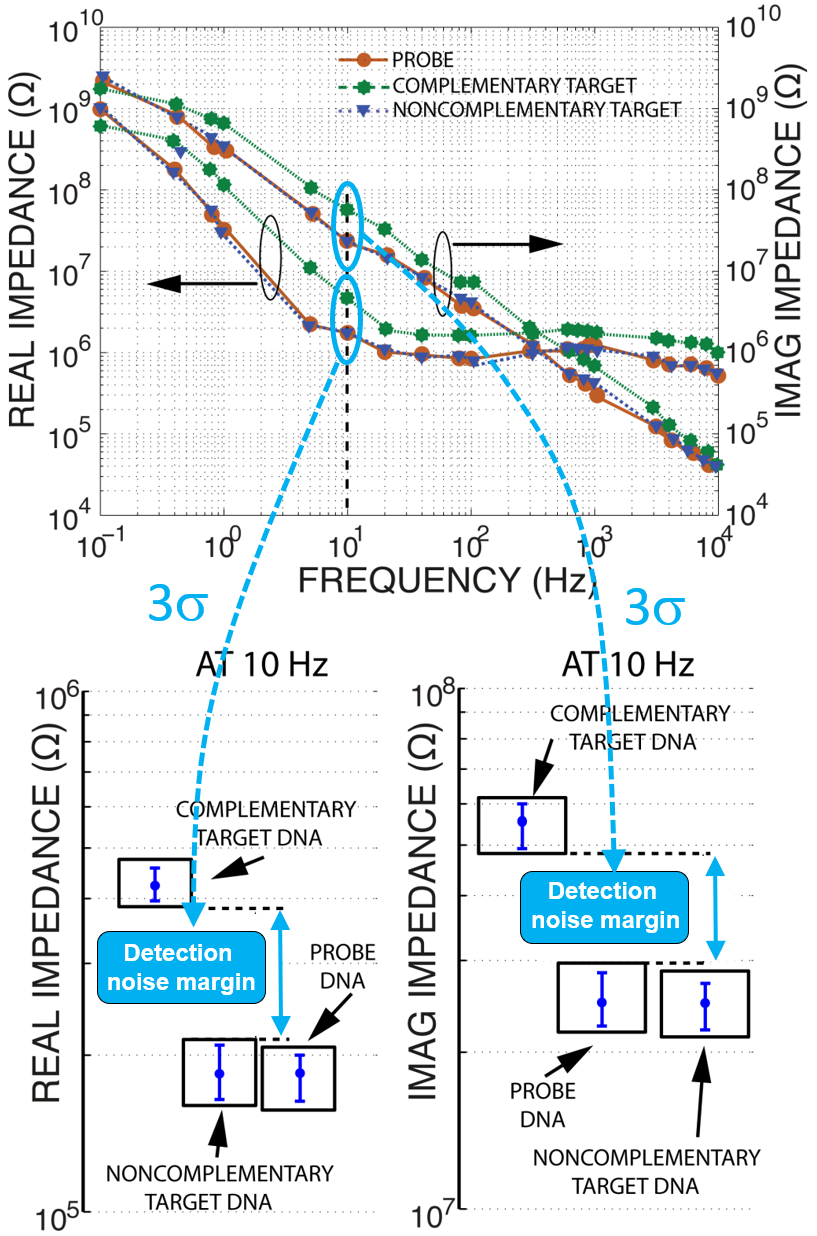}
	\caption{Experimental measurements of 5µM prostate cancer DNA using IS from 0.1Hz to 10kHz. The 3$\sigma$ error bars show 20 measurements from 3 chips.}
	\label{fig:exp_DNA_IS}
\end{figure}

Fig. \ref{fig:exp_DNA_CV} shows the experimental measurements of prostate cancer DNA in the CV mode. On-chip gold electrodes were used for the CV scans, which were performed in a 20µM potassium ferrocyanide solution.
Oxidation and reduction peaks are clearly visible in the CV scans using 5µM single stranded DNA modified electrodes. This is because of the negatively charged film created by thiolated DNA probes when they repell the negatively charged electrochemical reporter. Adding a 5µM non-complementary DNA target does not change the CV signal, which indicates that non-specific adsorption is negligible. On the other hand, adding a 5µM complementary target to the chip leads to the creation of double stranded DNA on the biosensing electrode, which indicates additional negative charge and elimination of ferrocyanide’s redox peak. The error bars (from 3 chips) in the figure show a detection noise margin of around 4nA.

Fig. \ref{fig:exp_DNA_IS} shows the experimental measurements of prostate cancer DNA using the IS mode. The IS frequency was from 0.1Hz to 10kHz. The recordings of 20µM potassium ferrocyanide in the cases of 5µM single stranded DNA, addition of 5µM non-complementary DNA, and addition of 5µM complementary DNA target are shown. The 3$\sigma$ error bars show 20 measurements from 3 chips. The impedance spectrum of DNA probe and non-complementary target are fairly similar as expected. The addition of the complementary DNA probe results in a change in the impedance spectrum.

Table~\ref{TABLE_COMP} provides a comparative analysis of the presented design and existing temperature regulated biochemical sensory microsystems. The presented work supports advantageous multi-modal DNA analysis with the most flexible temperature regulation scheme. Our design also achieves a compact channel area, thanks to the synergistic circuit sharing design.


\begin{table*}[!ht]
\centering
\caption{Comparative Analysis of Temperature-Regulated Biosensory Microsystems}
\label{TABLE_COMP}
\renewcommand{\arraystretch}{1.2} 
	\setlength{\tabcolsep}{9pt} 
\begin{tabu}{|[1pt]c|[1pt]c|[1pt]c|c|c|c|c|[1pt]c|[1pt]}
\Xhline{1pt}
                                                                        &                  & \begin{tabular}[c]{@{}c@{}}ISSCC\\ 2010 \cite{Wang}\end{tabular} & \begin{tabular}[c]{@{}c@{}}JSSC\\ 2017 \cite{Manickam2017}\end{tabular} & \begin{tabular}[c]{@{}c@{}}JSSC\\ 2019 \cite{Manickam2019}\end{tabular} & \begin{tabular}[c]{@{}c@{}}TBioCAS\\ 2020 \cite{Park2020}\end{tabular} & \begin{tabular}[c]{@{}c@{}}TBioCAS\\ 2020 \cite{Tedjo2020}\end{tabular} & This Work    \\ \Xhline{1pt}
\multirow{4}{*}{Modes}                                                  & CPA              & No                                                   & No                                                  & No                                                  & No                                                     & Yes                                                    & Yes          \\ \cline{2-8} 
                                                                        & CV               & No                                                   & No                                                  & Yes                                                 & No                                                     & Yes                                                    & Yes          \\ \cline{2-8} 
                                                                        & IS               & No                                                   & No                                                  & Yes                                                 & No                                                     & Yes                                                    & Yes          \\ \cline{2-8} 
                                                                        & Temp. Reg.       & Yes                                                  & Yes                                                 & Yes                                                 & Yes                                                    & Yes                                                    & Yes          \\ \Xhline{1pt}
\multirow{4}{*}{Process}                                                & CMOS Process     & 0.13µm                                               & 0.25µm                                              & 0.25µm                                              & 0.18µm                                                 & 0.6µm                                                  & 0.13µm       \\ \cline{2-8} 
                                                                        & Power            & 165mW                                                & 118mW                                               & 256mW                                               & 2.7mW                                                  & 1.242mW                                                & 0.35mW       \\ \cline{2-8} 
                                                                        & Supply Voltage   & 1.2V                                                 & 2.5V                                                & 2.5V                                                & 3.3V                                                   & 5V                                                     & 1.2V         \\ \cline{2-8} 
                                                                        & Chip Area        & 7.5mm$^2$                                               & 63mm$^2$                                               & 10.24mm$^2$                                            & 15mm$^2$                                                  & 361mm$^2$                                                 & 9mm$^2$         \\ \Xhline{1pt}
\multirow{8}{*}{Channel}                                                & Electrode Count  & 16                                                   & 1,024                                               & 1,024                                               & 1,225                                                  & 16,064                                                 & 600          \\ \cline{2-8} 
                                                                        & Electrode Type   & 2D                                                   & 2D                                                  & 2D                                                  & 2D                                                     & 2D                                                     & 3D           \\ \cline{2-8} 
                                                                        & Dynamic Range    & 55dB                                                 & 116dB                                               & 93dB                                                & N/A                                                    & N/A                                                    & 128dB        \\ \cline{2-8} 
                                                                        & Conversion Rate  & N/A                                                  & 50Hz                                                & 50Hz                                                & 17.06Hz                                                & 100kHz                                                 & 10kHz        \\ \cline{2-8} 
                                                                        & Sensitivity      & 0.3Hz                                                & 20fA                                                & 280fA                                               & N/A                                                    & 42.8pA                                                 & 8.6pA        \\ \cline{2-8} 
                                                                        & Signal Generator & Yes                                                  & No                                                  & Yes                                                 & Yes                                                    & Yes                                                    & Yes          \\ \cline{2-8} 
                                                                        & Signal Frequency & 1GHz                                                 & N/A                                                 & 50Hz                                                & N/A                                                    & N/A                                                    & 10kHz        \\ \cline{2-8} 
                                                                        & Wireless         & No                                                   & No                                                  & No                                                  & No                                                     & No                                                     & UWB          \\ \Xhline{1pt}
\multirow{5}{*}{\begin{tabular}[c]{@{}c@{}}Temp.\\ Regul.\end{tabular}} & Channel          & 1                                                    & 13                                                  & 13                                                  & 49                                                     & 2                                                      & 54           \\ \cline{2-8} 
                                                                        & Control Method   & PID                                                  & N/A                                                 & N/A                                                 & On/off                                                 & N/A                                                    & PID          \\ \cline{2-8} 
                                                                        & Implementation   & Off-chip                                             & Off-chip                                            & Off-chip                                            & On-chip                                                & Off-chip                                               & On-chip      \\ \cline{2-8} 
                                                                        & Signal Domain    & Analog                                               & N/A                                                 & N/A                                                 & Digital                                                & N/A                                                    & Mixed-signal \\ \cline{2-8} 
                                                                        & Accuracy         & 1$^{\circ}$C                                                    & 0.3$^{\circ}$C                                                 & N/A                                                 & 1$^{\circ}$C                                                      & 0.5$^{\circ}$C                                                    & ±0.5$^{\circ}$C          \\ \Xhline{1pt}
\end{tabu}
\end{table*}


\section{Conclusions}

A 9$\times$6 cell array (54-channel) mixed-signal CMOS temperature-regulated distributed DNA-sensing SoC was developed. The SoC reuses circuitry to perform CPA, CA, IS, and temperature regulation. The on-chip, in-channel heating and temperature sensing elements were implemented in standard CMOS without any post-processing. Local temperature can be regulated to within ±0.5°C of any desired point between 20$^{\circ}$C and 90$^{\circ}$C using PID control. The overall design achieved a very low power consumption at 42$\mu$W per channel from a 1.2V supply. Each channel occupies a silicon area of only 0.06mm$^{2}$. This design can be used in a wide range of applications where temperature regulation is desirable. The proposed circuit sharing methodology can also be beneficial for future highly integrated SoC designs.


\end{document}